\documentclass [smallextended] {svjour3}

\usepackage[T1]{fontenc}
\usepackage[latin9]{inputenc}
\usepackage[letterpaper]{geometry}
\usepackage{array}
\usepackage{float}
\usepackage{amsmath}
\usepackage{graphicx}
\usepackage{amssymb,cancel}
 \usepackage[colorlinks]{hyperref}

\def\beq{\begin{equation}}
\def\eeq{\end{equation}}
\newcommand{\bea}{\begin{eqnarray}}
\newcommand{\eea}{\end{eqnarray}}
\def\bmat{\begin{displaymath}}
\def\emat{\end{displaymath}}
\def\bear{\begin{eqnarray}}
\def\eear{\end{eqnarray}}
\usepackage{color}
\def\red{\color{black}}

\makeatletter

\providecommand{\tabularnewline}{\\}

\makeatother

\begin{document}

\title{A Non-Local Reality: Is there a Phase Uncertainty in Quantum Mechanics?}

\author{Elizabeth S. Gould \and Niayesh Afshordi}
\institute{Department of Physics and Astronomy,
University of Waterloo, 200 University Avenue West, Waterloo, ON, N2L 3G1, Canada \and 
Perimeter Institute for Theoretical Physics, 31 Caroline Street North, Waterloo, ON N2L 2Y5, Canada
\\Tel: +1 (519) 569-7600 Fax: +1 (519) 569-7611
\email{egould@pitp.ca}}


\maketitle

\begin{abstract}
{\red A century after the advent of Quantum Mechanics and General Relativity, both theories enjoy incredible empirical success, constituting the cornerstones of modern physics. Yet, paradoxically, they suffer from deep-rooted, so-far intractable, conflicts. Motivations for violations of the notion of relativistic locality include the Bell's inequalities for hidden variable theories,  the cosmological horizon problem, and Lorentz-violating approaches to quantum geometrodynamics, such as Horava-Lifshitz gravity.  Here, we explore a recent proposal for a ``real ensemble'' non-local description of quantum mechanics, in which ``particles'' can copy each others' observable values AND phases, independent of their spatial separation. We first specify the exact theory, ensuring that it is consistent and has (ordinary) quantum mechanics as a fixed point, where all particles with the same values for a given observable have the same phases. We then study the stability of this fixed point numerically, and analytically, for simple models. We provide evidence that most systems (in our study) are locally stable to small deviations from quantum mechanics, and furthermore, the phase variance per value of the observable, as well as systematic deviations from quantum mechanics, decay as $\sim$ (Energy$\times$Time)$^{-2n}$, where $n \geq 1$.  Interestingly, this convergence is controlled by the {\it absolute} value of energy (and not energy difference), suggesting a possible connection to gravitational physics. Finally, we discuss different issues related to this theory, as well as potential novel applications for the spectrum of primordial cosmological perturbations and the cosmological constant problem.}

\end{abstract}

\section{Introduction}

In his groundbreaking 1964 paper, John Bell demonstrated that the correlations predicted for quantum mechanical observables can never be fully replicated in {\it local hidden variable} theories \cite{Bell:1964kc}. The empirical success of quantum mechanics over the past 50 years would thus only be consistent with a hidden variable theory with superluminal (in fact, instantaneous) communication. However, experimental verification of Lorentz invariance (or equivalence principle) at extreme precisions (e.g. \cite{Liberati:2013xla}) has nearly vanquished any (empirical) motivation for such a possibility.  

Surprisingly though, a motivation for this possibility may come form cosmological observations: If quantum mechanics is given by a hidden variable theory, that theory
could have a non quantum mechanical extension with quantum mechanics
as an equilibrium fixed point. Then, non-local signalling \cite{valentini2002signal} may be possible {\it before} the universe enters
the quantum mechanical ``equilibrium'', and could be imprinted in the initial conditions of the universe. Therefore, one might be able to replicate the
correlations observed in the cosmic microwave background (CMB) (e.g. \cite{Ade:2013ktc})  using the non-local signalling 
in the pre-quantum mechanical universe \cite{valentini2002signal}, as they cannot otherwise be explained in the standard Big Bang theory (so-called horizon problem). 

{\red Another motivation for instantaneous signalling comes from a power-counting renormalizable approach to 3+1d geometrodynamics, known as Horava-Lifshitz gravity \cite{Horava:2009uw} which, violates foliation invariance (and thus relativistic locality) of general relativity. At high energies, the speed of propagation for excitations in this theory approaches infinity, amounting to non-local (though causal) instantaneous signalling. } 

A real ensemble model is a 
non-local
hidden variable theory, as an alternative to quantum mechanics, and so can potentially possess this type of non-local signalling.
In order for this theory to be viable, however, quantum mechanics must be
an attractor, such that the non-quantum mechanical theory becomes
quantum mechanics at later times, consistent with present-day experiments. Valentini shows this is the case
for Bohmian mechanics \cite{PhysRev.85.166,PhysRev.85.180} using a coarse-grained H-theorem as in statistical
mechanics, so that one can see there are regimes which approach quantum
equilibrium \cite{quant-ph/0104067}. The proof applies to the coarse-grained
case for a model with a real wavefunction and independent probability.

In this paper, we focus on a different hidden variable model,  the so-called "real ensemble" model
recently introduced by Smolin \cite{smolin2011real}. The model focuses on an object described in quantum mechanics by a wavefunction $\left|\psi\right>$, 
which could be a particle, field, composite system of many particles, etc. We will refer to this as the system. It puts all systems that would be in
the same state in quantum mechanics into an ensemble. One would then examine this system in a given eigenbasis for the observable of interest. 
There are two beables\footnote{Beables generally refer to properties innate to the system rather than determined by observation. This term is used in 
hidden variable theories to give a name to the properties of the system which actually exists and are subject to the laws of evolution of the system. It 
typically includes both the observables and the hidden variables.} for each such system - the value of the observable in 
question and the phase of the component of the wavefunction in the eigenstate 
corresponding to the observable value would possess in quantum mechanics. All these systems are spread out across the universe, 
interacting non-locally as governed by the rules of interaction of members of the ensemble. The systems
then evolve according to two rules: {\it i)} The continuous evolution rule
states that the phase of the systems evolves according to
some equation, and {\it ii)} The copy rule states that there is a finite probability
for one system in the ensemble to change its {\red observable and phase values} to match those
of another system's. This then can be used to determine the
evolution of the probability that a system would be in a given state
as well as the phase associated with that state.

The model is called ``real ensemble'' since all the systems can exist in the same universe and their 
Hamiltonians which determine their local interactions can, in principle, include the influence of other members of the ensemble. This conception is in contrast to the ``many-worlds interpretation", where the ensemble is effectively over multiple parallel "universes" (or rather separate portions of the wavefunction), but for what would be considered the same system in each.

Since in Smolin's real ensemble model the phases (of the wave function) are the additional hidden variables, the non-equilibrium behaviour is different. 
While we do not have a coarse-grained H-theorem, we have examined the convergence numerically and analytically to determine not only that there are 
regimes for which the model converges, but also the rate of convergence is dependent on {\it absolute energy}.
%

{\red The outline of the paper is as follows: We begin in Sec. \ref{framework} by developing a non-equilibrium real ensemble model,
which admits quantum mechanics (Eqs. \ref{eq:origphidot}-\ref{eq:origrhodot}) as an equilibrium limit. We also study the dynamics close to the equilibrium.  
In Sec. \ref{numeric}, we study the stability properties of the non-equilibrium model numerically for simple spin$-\frac{1}{2}$ systems, and in particular 
examine the stable parameter-space of the theory. In Sec. \ref{near}, we provide an analytic perturbative study of near-equilibrium behaviour, which is 
roughly consistent with the numerical results. Finally, Sec. \ref{discuss} puts the real ensemble framework, along with our findings, into some physical 
context, and Sec. \ref{conclude} concludes the paper.}

\section{ Real Ensemble Theory: the framework }\label{framework}

We begin by introducing the equilibrium real ensemble model which reproduces quantum mechanics. In the case of equilibrium, there is one phase per observable value. 
For this case, the number of systems, $N$, is sufficiently large such that there is a large enough number of systems
per distinct sets of beables (phases AND observable values), so that they can be treated as continuous numbers. 
We define the probability for a system to have a given value $a$ for the observable of interest as $\rho_a(t)$. 
The phase associated with this observable value is defined as $\phi_a(t)$. The equations which reproduce quantum mechanics are then \cite{smolin2011real}:
\begin{equation}
\dot{\phi}_a\left(t\right)=
\underset{b}{\sum}\sqrt{\frac{\rho_b\left(t\right)}{\rho_a\left(t\right)}}\, R_{a b}\cos\left[\phi_a\left(t\right)-\phi_b\left(t\right)+\beta_{a b}\right],\label{eq:origphidot}\end{equation}

\begin{equation}
\dot{\rho}_a\left(t\right)=\underset{b}{\sum}2\sqrt{\rho_a\left(t\right)\rho_b\left(t\right)}R_{ab}\sin\left[\phi_a\left(t\right)-\phi_b\left(t\right)+\beta_{ab}\right],\label{eq:origrhodot}\end{equation}
where the evolution $\dot{\phi}$ comes from the continuous evolution rule and the evolution $\dot{\rho}$ from the copy rule 
\footnote{In order to match quantum mechanics, one needs to include a factor of 2 that was missing from Equation 30 from 
\cite{smolin2011real}. This correction, however, will not effect any later conclusions, and it was fixed in the published version. 
The other change in this from Smolin's notation is the replacement of $\delta$ with $\beta$ to prevent confusion with the Kronecker $\delta$-function.}. 
$R$ and $\beta$ in this equation are real, non-negative numbers defined by the particular system.
This case is identical to the quantum mechanical system with the wavefunction given by
\begin{equation}
\left|\Psi\right>=\underset{a}{\sum}\sqrt{\rho_a\left(t\right)}\, e^{-i\phi_a}\left|a\right>,
\end{equation}
and Hamiltonian given by
\begin{equation}
H=\hbar\underset{a,b}{\sum}R_{a b}e^{i\beta_{a b}}\left|a\right>\left<b\right|.
\label{eq:hamilorig}\end{equation}
Equation \ref{eq:origphidot} is equivalent to equation 31 in \cite{smolin2011real}, equation \ref{eq:origrhodot} to equation 30.

In order to determine the non-equilibrium forms of the equations,
we need to remember that the calculation of the two Equations \ref{eq:origphidot}
and \ref{eq:origrhodot} starts with the assumption that $\phi$ is
a function of the value of the observable rather than the system, defined as
the equilibrium condition. The goal is to extrapolate these equations
to non-equilibrium equations, one which has $\phi$ different for
different systems even if the value of the observable is the same, using the
known equilibrium equations. $\dot{\phi}$ and $\dot{\rho}$ must
reduce to those shown in Equations \ref{eq:origphidot} and \ref{eq:origrhodot}
when there is only one potential phase per {\red value of the observable}, as this will reproduce
quantum mechanics when the function achieves equilibrium. We will
retain the large $N$ limit assumption.

%
%
%

\subsection{Allowing for Multiple Phases per Value of Observable}

\begin{figure}[H]
\begin{center}
\def\svgwidth{0.50\paperwidth}

\input{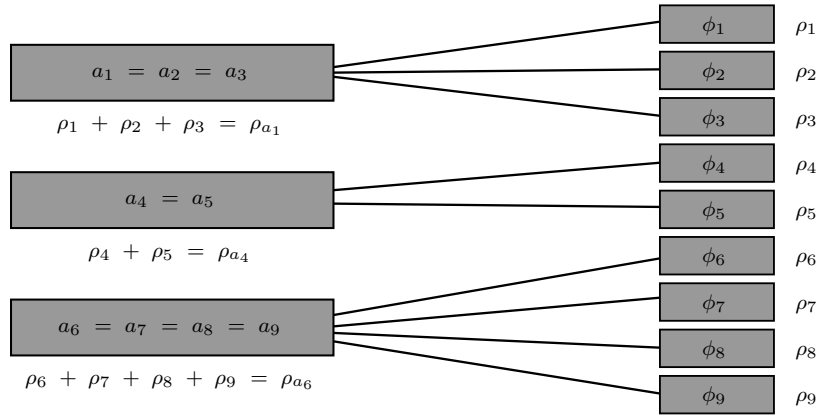}
\end{center}

\caption{\label{fig:ExpDblSm} A pictorial depiction of an out of equilibrium real-ensemble system: Each value of observable, $a_i$, has multiple potential values of phase, $\phi_i$. $\rho_i$ 
on the right is the probability for a particle to have both an 
observable value $a_i$ and a phase $\phi_i$. In contrast, $\rho_{a_i}$ on the left is the probability 
of having $a_i$ but with any phase. The sum of a given quantity over all $i$ can either be determined by taking the sum over all indices $i$, or by taking the sum over all observable values 
and all phases with non-zero probability for each observable value. In equilibrium, there is only one $\phi$ for each $a_i$, i.e. $\phi_i=\phi_{a_i}$.}

\end{figure}

The number of cases for a given value of the observable can be represented as a
sum of the number of systems with a given set of beables for all
beable sets with the given observable value $a$. Working in
probabilities since the number of systems $N$ is large, this can
be replaced by a weight function $w$ which determines the probability
of the system with a given value of an observable having a given
phase. As a probability function, 
\begin{equation}
1=\underset{\phi}{\sum}w\left(a,\phi,t\right)
\end{equation}
for every $a$ and $t$. We will label a system type - all systems which share phase 
and observable value - by the labels $i$, $j$, $k$, etc., rewriting $a$ as $a_i$, $\rho_a$ as $\rho_{a_i}$, and $\phi_a$ as $\phi_{a_i}$. 
In the non-equilibrium case, $\rho_{a_i}$ becomes $\rho_i$ and $\phi_{a_i}$ becomes $\phi_i$. Note that $\rho_{a_i}$ still has meaning
out of equilibrium as $\sum_j \rho_j \delta_{a_i a_j}$. $\delta$ here is the Kronecker delta. A depiction of this expansion of 
the number of distinct pairs of beables and the labeling is shown in Figure \ref{fig:ExpDblSm}. 
Using $w$, $\rho_i = \rho_{a_i} w_i$.
Since $a_{j}$ is no longer the only quantity
the terms in the sum are dependent on, $\sum_{a_{j}} \cdots$ becomes 
\begin{equation}
\sum_{a_{j}}\sum_{\phi_{j}}w\left(a_{j},\phi_{j},t\right) \cdots  \equiv \sum_j w_j~ \cdots ,
\end{equation}
where $\cdots$ is whichever expression the sum is taken over in the equations of evolution. 
The process of converting from a double sum to a single sum is a simplification of the notation determined by recognizing there are multiple values of $j$ 
with the same $a_j$, each with different $\phi_j$. This can be seen in Figure \ref{fig:ExpDblSm} where it is a sum over three $a_j$, and for 
each $a_j$, a sum over two to four $\phi_j$ converted to a sum over nine values of $j$.

However, there is a less obvious place in which this weight function
is needed. Underneath the square roots $\rho_{a_{j}}$
may not be independent of phase anymore. It is uncertain if it only
depends on the probability of a value of the observable or on the probability of
a given beable pair. This is dealt with by adding in an arbitrary real function $F\left(a,\phi_{1},\phi_{2}\right)$
for which $F\left(a,\phi_{1},\phi_{1}\right)=1$ and $F\left(a,\phi_{1},\phi_{2}\right)\geq0$
everywhere. This function determines the contribution to the density
functions under the square root of systems with different phases
but the same value of the observable. 
{\red We must then replace} $\rho_a$
by 
\begin{alignat}{2}
\tilde{\rho}_{a_i} & \equiv & \rho_{a_i}\underset{j}{\sum}w_j \delta_{a_i a_j} F\left(a_i,\phi_{i},\phi_{j}\right)\\
 & \equiv & \underset{j}{\sum} \rho_j \delta_{a_i a_j} F\left(a_i,\phi_{i},\phi_{j}\right).
\label{eq:rhotildedef}
\end{alignat}
$F$ must be periodic with respect to $\phi_{1}-\phi_{2}$, but might
have additional dependence on $a$ and $\phi_{1}$.
Since this is an extenstion beyond quantum mechanics, we have no guide to determine what form this extension should take,
but we can examine several different models to test for convergence. We will minimize the symmetries
broken when leaving equillibrium since there is no reason to suspect that they are broken and this reduces the potential forms of $F$.
A $\phi_{1}$
dependence could introduce a dependence on absolute phase, while a
dependence on $a$ may remove certain quantum mechanical symmetries
from some systems when leaving equilibrium. These additional dependencies
will not be considered here since there is no reason to believe that
such dependencies exist nor any need to consider them. In addition,
only the case for which $F\left(\phi_{1}-\phi_{2}\right)=F\left(\phi_{2}-\phi_{1}\right)$
will be considered. If this does not hold, time reversal invariance
will no longer hold: Since under time reversal in this model, $\phi\rightarrow-\phi$,
$t\rightarrow-t$, $\rho\rightarrow\rho$, $R\rightarrow R$, and
$\delta\rightarrow-\delta$\cite{smolin2011real}, applying time reversal
to Equations \ref{eq:phidotnew} and \ref{eq:rhodotnew} (the final
equations, later) will only give $\dot{\rho}\rightarrow-\dot{\rho}$
and $\dot{\phi}\rightarrow\dot{\phi}$ for all $a$, $\phi$, and $\rho\left(a,\phi,t\right)$
if $F\left(\phi_{1}-\phi_{2}\right)=F\left(\phi_{2}-\phi_{1}\right)$.

{\red We further exclude the possibility that $F$ is} $1$ when $\phi_{1}=\phi_{2}$
and $0$ otherwise, because then transferring a large number of systems
from $\phi_1\neq\phi_2$ to $\phi_1=\phi_2$ continuously will result
in a discontinuity in the function $\dot{\phi}$ and $\dot{\rho}$
from the discontinuity in $\rho\left(a,t\right)\sum w\left(a,\phi_{j},t\right)F\left(a,\phi_{i},\phi_{j}\right)$.
Other functions which similarly produce a discontinuity when going
to equilibrium will be problematic.

The resulting evolution equations are \begin{equation}
\dot{\phi}_i\left(t\right)=\underset{j}{\sum}w_j\left(t\right)
\sqrt{\frac{\rho_{a_{j}}\left(t\right)\underset{k}{\sum}w_k\left(t\right) \delta_{a_j a_k} F\left(\phi_{j}-\phi_{k}\right)}{\rho\left(a_{i},t\right)\underset{k}{\sum}w_k\left(t\right) \delta_{a_i a_k} F\left(\phi_{i}-\phi_{k}\right)}}
\times R_{a_{i} a_{j}}\cos\left[\phi_{i}\left(t\right)-\phi_{j}\left(t\right)+\beta_{a_{i} a_{j}}\right]\quad\textrm{and}\label{eq:phidotnew0}\end{equation}
\begin{multline}
\dot{\rho}_i\left(t\right)=w_i\left(t\right)\underset{j}{\sum}w_j\left(t\right)
\sqrt{{\scriptstyle \rho_{a_{i}}\left(t\right)\left[\underset{k}{\sum}w_k\left(t\right) \delta_{a_i a_k} F\left(\phi_{i}-\phi_{k}\right)\right]\rho_{a_{j}}\left(t\right)\left[\underset{k}{\sum}w_k\left(t\right) \delta_{a_j a_k} F\left(\phi_{j}-\phi_{k}\right)\right]}}\\
\times2R_{a_{i} a_{j}}\sin\left[\phi_{i}\left(t\right)-\phi_{j}\left(t\right)+\beta_{a_{i} a_{j}}\right].\label{eq:rhodotnew0}\end{multline}

%
%
Equivalently, we can define the probability for a given value of the observable and phase,  $\rho_i \equiv \rho_{a_i} w_i$, yielding:
\begin{equation}
\dot{\phi}_i = \underset{j}{\sum}
\frac{\rho_j}{\underset{k}{\sum}\rho_k \delta_{a_j a_k}}
\sqrt{\frac{\tilde{\rho}_{a_j}}{\tilde{\rho}_{a_i}}}
\times R_{a_{i} a_{j}}\cos\left[\phi_{i}-\phi_{j}+\beta_{a_{i} a_{j}}\right],\label{eq:phidotnew}
\end{equation}
\begin{equation}
\dot{\rho}_i = 
\frac{\rho_i}{\underset{k}{\sum}\rho_k \delta_{a_i a_k}}
\underset{j}{\sum}
\frac{\rho_j}{\underset{k}{\sum}\rho_k \delta_{a_j a_k}}
\sqrt{{\scriptstyle \tilde{\rho}_{a_i}\tilde{\rho}_{a_j}}}
\times2R_{a_{i} a_{j}}\sin\left[\phi_{i}-\phi_{j}+\beta_{a_{i} a_{j}}\right].\label{eq:rhodotnew}
\end{equation}
Here, as in equation \ref{eq:rhotildedef}, 
\begin{equation}
\tilde{\rho}_{a_i} \equiv \underset{k}{\sum}\rho_k \delta_{a_j a_k} F\left(\phi_{j}-\phi_{k}\right).
\end{equation}

First note that $R_{a b}$ and $\beta_{a b}$ are symmetric and anti-symmetric, respectively, due to the hermiticity of the Hamiltonian. Therefore, summing 
Eq. (\ref{eq:rhodotnew}) above over $a_i$ and $\phi_i$, we end up with terms that are completely symmetric under $i \leftrightarrow j$, with the exception of 
the sine factor, which is anti-symmetric. As a result, the sum vanishes and thus the total probability remains conserved. 

\subsection{Other Changes When Leaving Equilibrium}

While they will not be covered here, there are a few other additions
to this equation which might be examined. The first is the addition
of terms which go to zero for the large $N$ equilibrium case. There
is nothing to indicate what form these terms might take, except that
they must go to zero for either the large $N$ limit or the quantum
mechanical limit, and they can't cause any probabilities to not behave
as probabilities. Other possibilities include adding a $\phi$ dependence
into $R$ and $\beta$. This would have to take a form in which the
quantum mechanical limit still holds, say by taking the difference
between the mean phase and $\phi$, or the spread of phases as influencing
$R$.

\section{Discrete Systems: Spin-$\frac{1}{2}$  with Finite Number of
Phases per Value of Observable}\label{numeric}

In this section, we numerically study the evolution of simplest possible systems, i.e.  spin-$\frac{1}{2}$'s or qubits, and investigate whether/how their evolution 
approaches the quantum mechanical equilibrium.    

\subsection{The Model}

This case involves two possible values of $a$, spin up or spin down,
simplifying the analysis since there are only two potential values of the observable to sum over.
The input to the algorithm will determine the $R$ and $\beta$ parameters
by using the sum of Pauli matrices to define a Hamiltonian, $H=c_{t}I+c_{x}\sigma_{x}+c_{y}\sigma_{y}+c_{z}\sigma_{z}$.
The $R$ and $\beta$ functions are determined by the Hamiltonian in equation \ref{eq:hamilorig}
using \begin{equation}
R_{1 2}=R_{2 1}=\sqrt{c_{x}^{2}+c_{y}^{2}};\; R_{1 1}=\left|c_{t}+c_{z}\right|;\; R_{2 2}=\left|c_{t}-c_{z}\right|\end{equation}
\begin{equation}
\beta_{1 2}=-\beta_{2 1}=\arg\left(c_{x}-ic_{y}\right);\;\beta_{1 1}=\arg\left(c_{t}+c_{z}\right);\;\beta_{2 2}=\arg\left(c_{t}-c_{z}\right).\end{equation}

The function $F$ will have one of a few predefined values constructed as sample functions using the limitations in previous section. The first
possibility is a constant F, $F\left(\Delta\phi\right)=1$. This corresponds
to the case when the densities under the square roots are independent
of phase.

The smooth cosine type function is a simple solution to account for
the phase differences as being important while still creating an influence
from the variation in the phases of the different systems. This
would appear as \begin{equation}
F\left(\Delta\phi\right)\equiv \frac{1}{2}+\frac{1}{2}\cos\left(\Delta\phi\right)=\cos^{2}\left(\frac{\Delta\phi}{2}\right).\label{eq:cosF}\end{equation}

{\red This definition can be generalized to model a sharper dependence on phase difference.  
One such function is 
\beq
F_c\left(\Delta\phi\right) \equiv \cos^{2}\left(\frac{c ~\Delta\phi}{2}\right) \Theta[\cos(\Delta\phi) -\cos(\pi/c)], \label{eq:spikeF}
\eeq
for a constant $c$, where $\Theta[x]$ is the step function that vanishes for $x<0$ and is $1$ otherwise.  We note that $F_1=F$ defined in Eq. (\ref{eq:cosF}).  Moreover, for 
completeness, we define $F_0(\Delta\phi) \equiv1$. 

For our numerical studies  below, unless noted otherwise, we focus on three possibilities: $F(\Delta\phi)=F_0, F_1,$ and $F_{100}$. }

\subsection{Results and Plots}

{\red We numerically evolve Eqs. (\ref{eq:phidotnew}-\ref{eq:rhodotnew}) with either two or three different phases per each
potential value of $s_z$. The results can be seen in Table \ref{tab:Plots-of-Equilibrium}, as well as in the appendix.} 
In the plots, spin up is represented by the blue colours (dotted black line, dashed blue line,
thin purple line), while spin down by the red colours (wide dashed red line, dot-dashed orange line, solid pink line).
The plots in Table \ref{tab:Plots-of-Equilibrium} are representative
of all the cases run with similar {\red initial} differences in the phases
of the same value of $s_z$. In each case, the sum of the probabilities of measuring a
specific value of $s_z$ follow an evolution indistinguishable within the
expected error from the quantum mechanical case.

\begin{table*}
\caption{\label{tab:Plots-of-Equilibrium}Plots of the evolution of spin-$\frac{1}{2}$
systems in the non-equilibrium real ensemble model. Each case has
three different phases for each of the two potential values of $s_z$. The initial
conditions are $\rho\left(0\right)=\left\{ \left\{ 0.16,0.08,0.06\right\} ,\left\{ 0.23,0.3,0.17\right\} \right\} $
and $\phi\left(0\right)=\left\{ \left\{ 0,0.001\pi,0.002\pi\right\} ,\left\{ \frac{\pi}{2}+0.001\pi,\frac{\pi}{2},\frac{\pi}{2}+0.0005\pi\right\} \right\} $.
The Hamiltonian is $H=\omega_{0}\hbar\left(2\sigma_{z}\right)$. $\hbar\equiv1$
and $\omega_{0}$ determines the units for $t$. From top to bottom,
the functions $F$ within equations \ref{eq:phidotnew} and \ref{eq:rhodotnew}
for the plots are $F=1$, $F=\cos^{2}\left(\frac{\Delta\phi}{2}\right)$,
and $F$ given by equation \ref{eq:spikeF} with $c=100$.}

\begin{tabular}{|c|c|c|}
\hline 
 & Probability vs Time & Phase Difference vs Time\tabularnewline
\hline
\hline 
\includegraphics[width=0.0173\paperwidth]{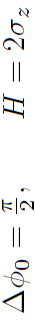} & \includegraphics[width=0.32\paperwidth]{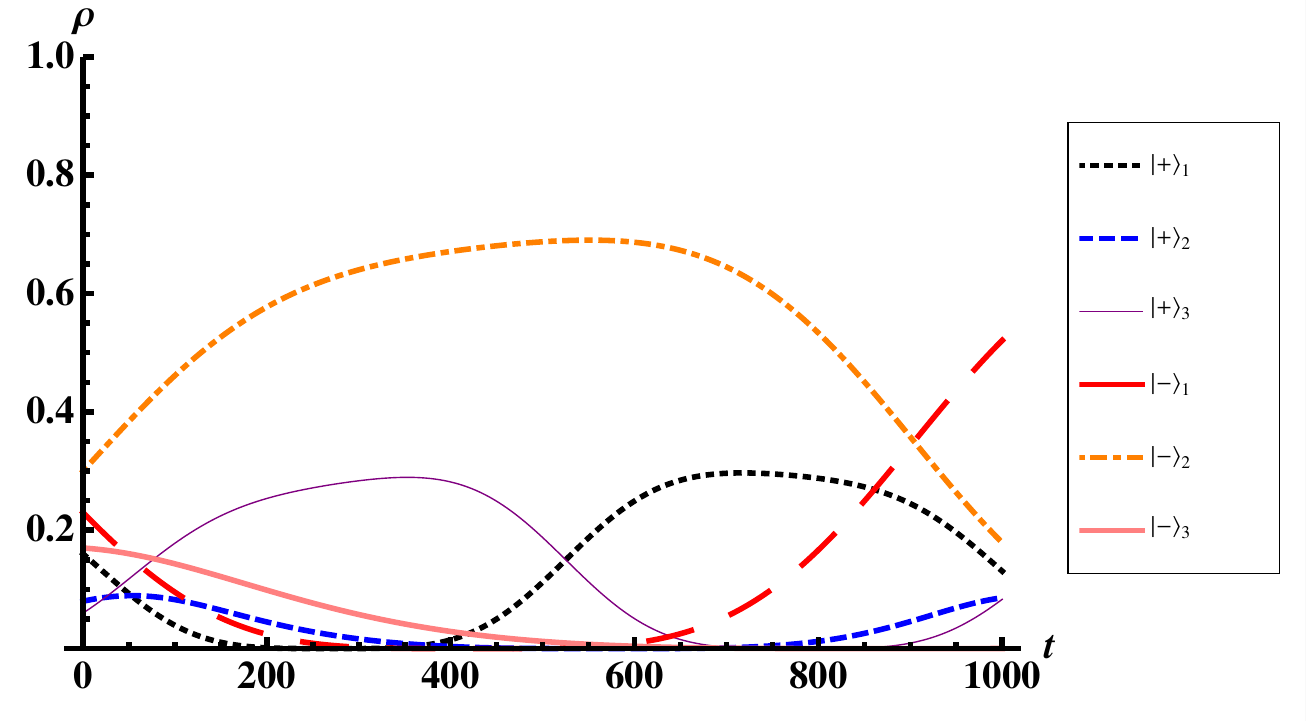} & \includegraphics[width=0.26\paperwidth]{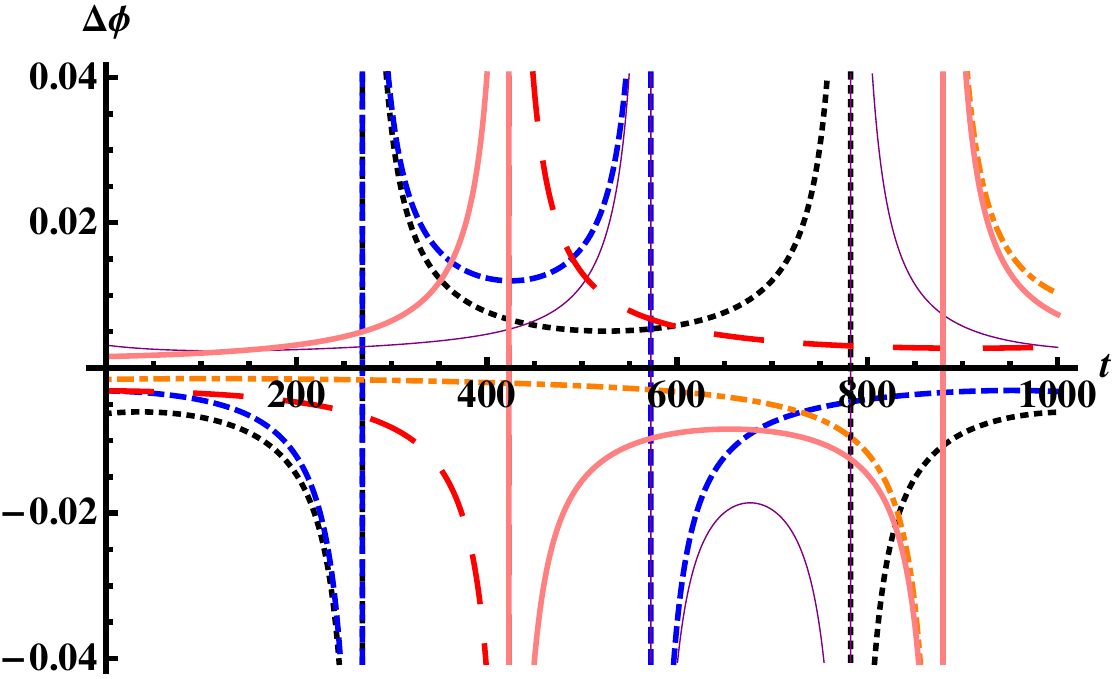}\tabularnewline
\hline 
\includegraphics[width=0.0173\paperwidth]{ALzz2} & \includegraphics[width=0.32\paperwidth]{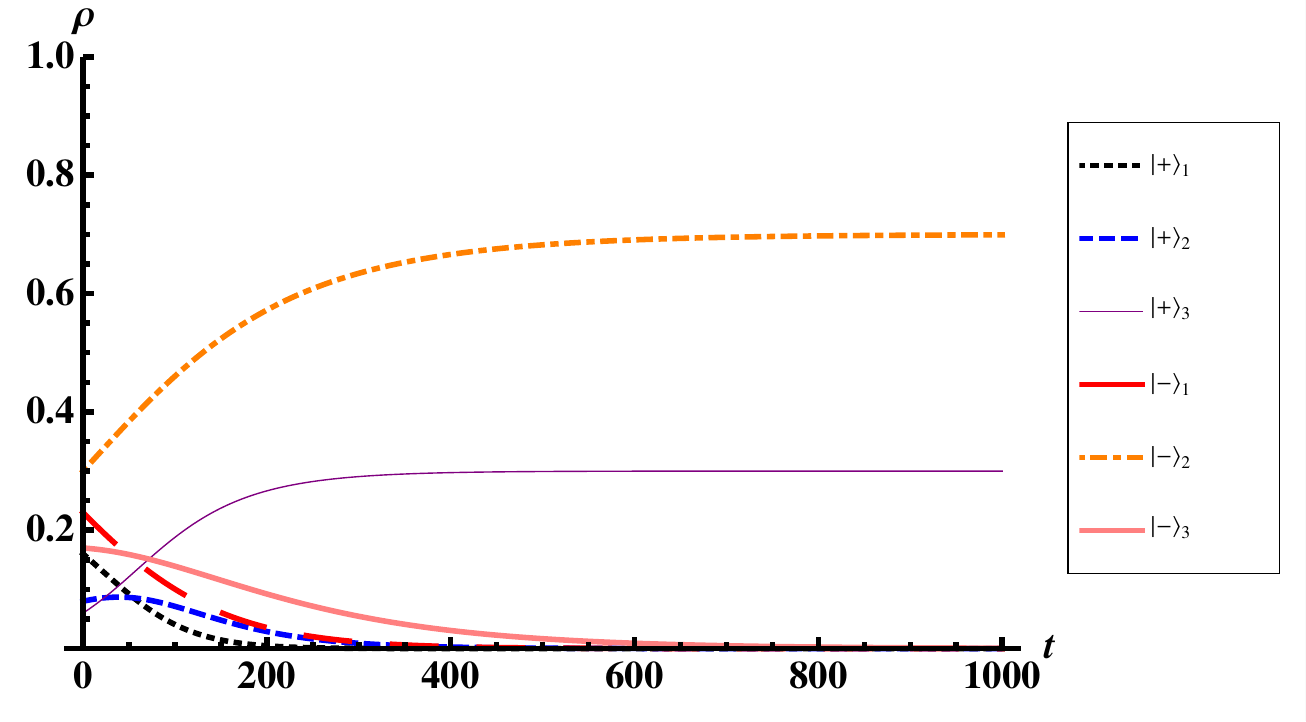} & \includegraphics[width=0.26\paperwidth]{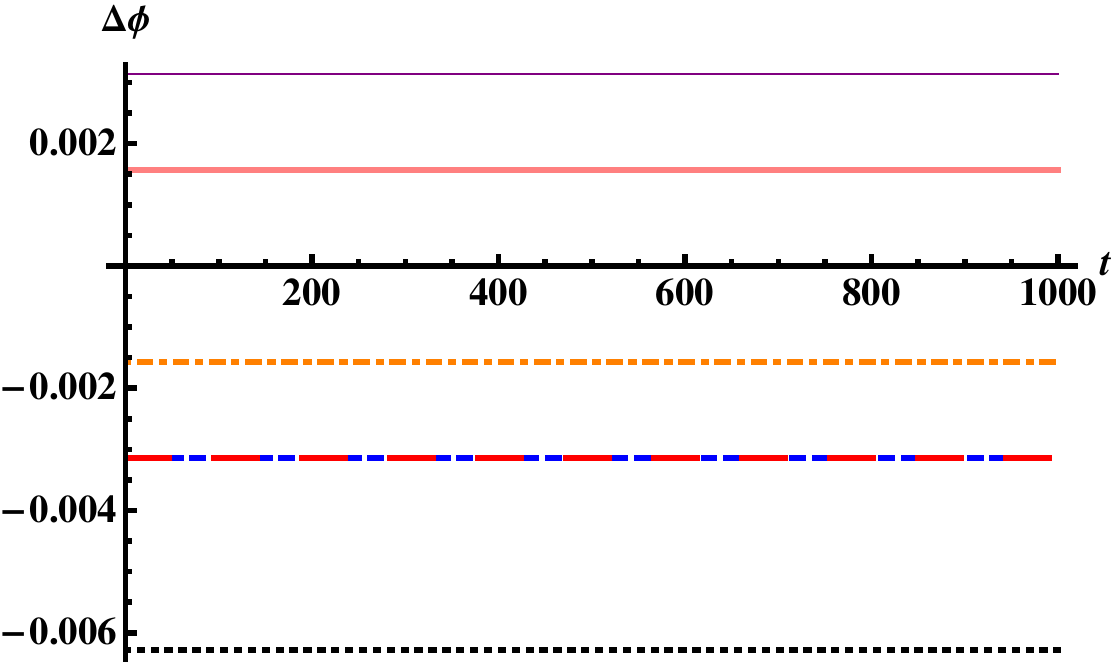}\tabularnewline
\hline 
\includegraphics[width=0.0173\paperwidth]{ALzz2} & \includegraphics[width=0.32\paperwidth]{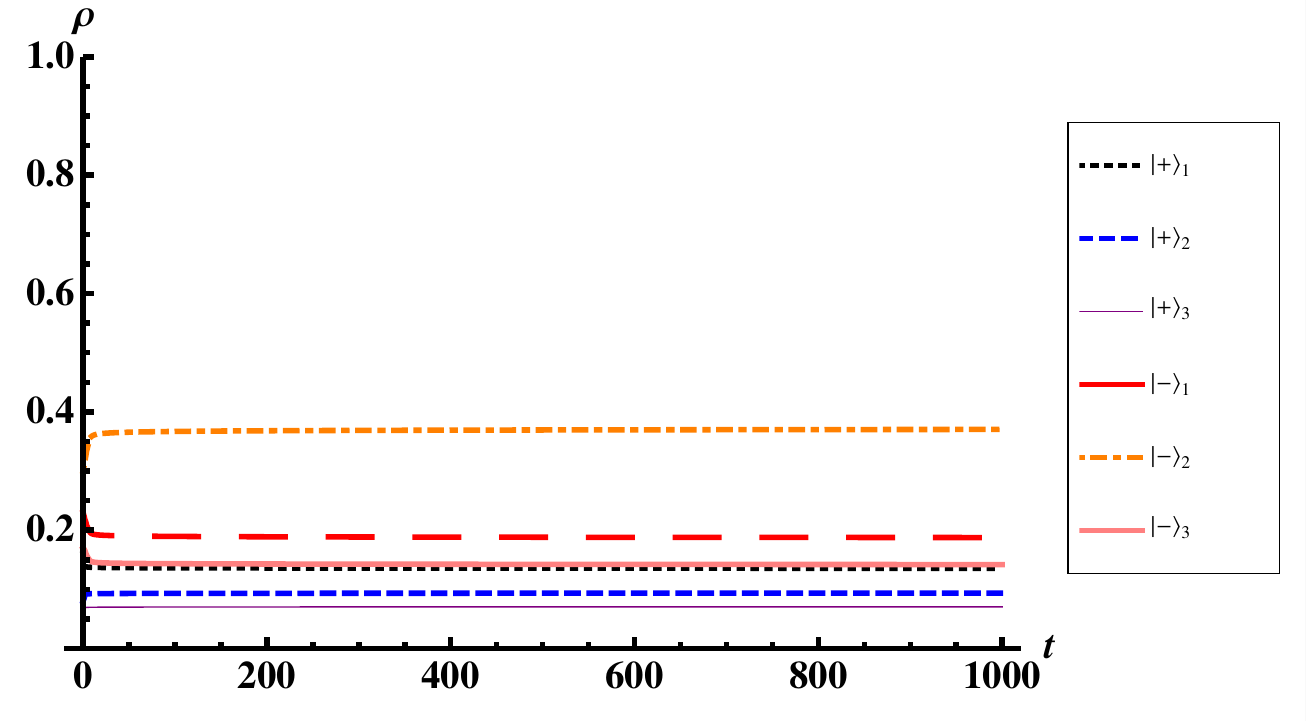} & \includegraphics[width=0.26\paperwidth]{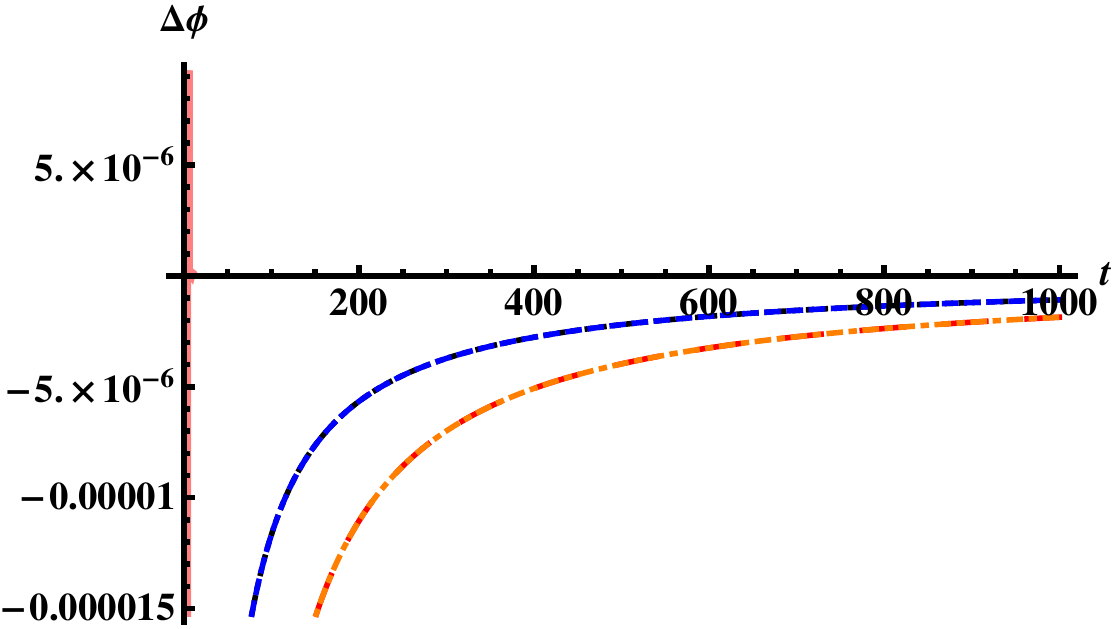}\tabularnewline
\hline
\end{tabular}
\end{table*}

For the two cases of functions which are flat or close to flat within
this region, they start with the same behaviour, but the kernel that slightly
deviates from flat  ({\red i.e. $F_1$}) is stable and fully approaches equilibrium,
while the other case, {\red $F_0$} does not and diverges from equilibrium. The third
case, $F_{100}$, which {\red which has the narrowest domain}
causes the system to approach equilibrium in a different way: Instead
of the probability for certain pairs of beables to be present dropping to zero, the difference
between the phases of the beable pairs with the same observable value goes to zero.

One {\red subtlety} to note for the $F_0$ case is that the values given do go
below the expected error threshold for the  {\red numerical} simulation. This, 
however, does not nullify the results which are seen - the fact that this
is unstable remains, but the values after the point in which the model 
decreases below the error threshold can not be trusted. 

For both the $F_0$ and $F_1$ cases, the very low probabilities could approach 
the levels in which the discreteness of the set are relevant and the large 
$N$ approximation breaks down for that particular beable pair. This possibility 
could alter the evolution of these beable pairs, and more interestingly potentially
cause one of the potential pairs of beables to completely empty, removing it from the ensemble. 
This could possibly cause the $F_0$ case to become quantum mechanical, despite the 
instability, or the $F_1$ case to become exactly equilibrium.

For the evolution with greater initial difference in the phases, the cases
with a phase difference of 0.1 still go to equilibrium for those $F$ functions
which approached equilibrium for smaller initial angle difference, while the 
cases with a relatively large phase difference do not. At the start, however, 
the evolution of the sum of all probabilities of measuring one of the two values of $s_z$ is not necessarily
quantum mechanical for even the moderate phase difference. Sample plots for the 
$F_1$ and $F_{100}$ cases are given in Tables \ref{tab:Plots-of-Equilibrium-1-1} and 
\ref{tab:Plots-of-Equilibrium-1} in the Appendix.

\subsection{\red Rate of Convergence}

In an examination of the small phase difference for which shape converges
fastest, one must first develop a measure for the rate of convergence.
A simple such definition would be to use the standard deviation, as
it approaches zero for either method of approaching equilibrium. Table
\ref{tab:Plots-of-Equilibrium-2} shows the distance from quantum
mechanics using this measure vs. time. Note that the $F_{100}$ case here is
log-log, showing an {\red approximate $\sim t^{-1}$ decay,} while the $F_1$ case is logarithmic only
on the y-axis, showing an exponential decay/convergence.

It can be seen that with this measure of convergence, the $F$ function
given by a spike ($c=100$) converges  to quantum mechanics faster within the given 
time, than that given by a cosine function ($c=1$). This indicates a faster
rate of convergence at early times, for a sharper $F$-kernel. The flat $F$ function ($c=0$) approaches 
equilibrium in a similar rate as the cosine function, until it jumps away 
from equilibrium.

\begin{figure}[H]
\includegraphics[width=0.70\paperwidth]{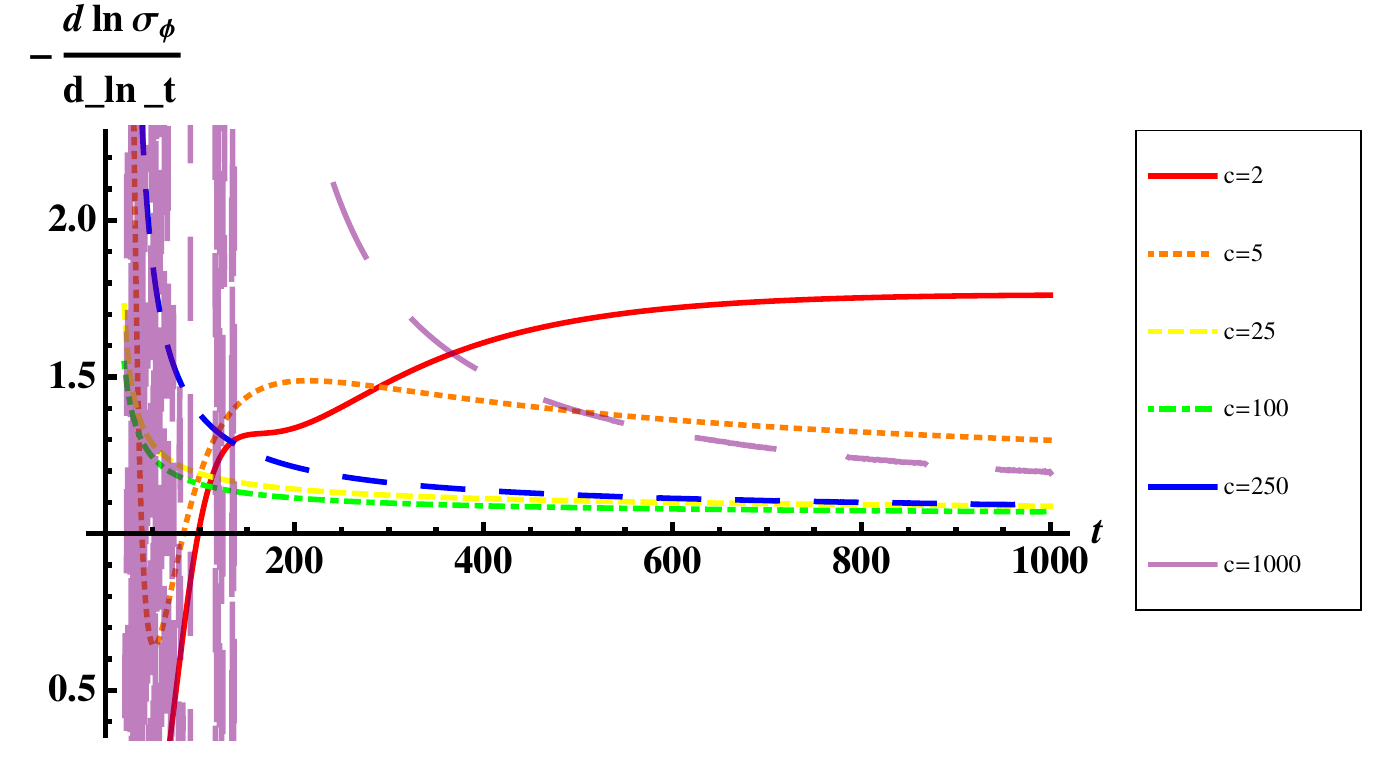}

\caption{\label{fig:MasterPlot2} This plot shows the asymptotic behaviour of the rate of convergence, which we quantify as the logarithmic time derivative of the standard deviation of the phase: $n= \frac{d \ln\sigma_\phi}{d\ln t}$. Shown is the plot of $-n$ vs $t$. The colours red (solid line), orange (dotted line), yellow (dashed line just above the green/dot-dashed line), green (dot dashed line), blue (wide dashed line) and purple (dashing wider than blue case, consists of spike behaviour before $t=200$) correspond to $c=2, 5, 25, 100, 250$ and $1000$  respectively. We see that convergence is always faster than $\sigma_\phi \propto t^{-1}$. The case $c=1$ is not plotted here, as it decays exponentially $ n\rightarrow -\infty$.}

\end{figure}

For the case of $H=2\sigma_{z}$, Figure \ref{fig:MasterPlot} shows if the distribution converges to quantum mechanics for a given width c of the F function in equation \ref{eq:spikeF} 
and initial phase separation $\Delta\phi_{0}$ for $\rho\left(0\right)=\left\{ \left\{ 0.16,0.08,0.06\right\} ,\left\{ 0.23,0.3,0.17\right\} \right\}$  
and $\phi\left(0\right)=\left\{ \left\{ 0,\alpha,2\alpha\right\} ,\left\{ \frac{\pi}{2}+\alpha,\frac{\pi}{2},\frac{\pi}{2}+\frac{\alpha}{2}\right\} \right\}$. The case for $c=0$, 
equivalent to $F=1$, is not shown on the plot, and does not converge. The convergence can be seen for an initial phase separation less than $\frac{\pi}{10}$, while for values 
above $\Delta\phi_{0} \sim 1$, the system {\it does not} approach quantum mechanics. Convergence here is defined as having the variance decrease faster than approximately $t^{-0.2}$ 
by $t=1000$, however most cases are obvious with a variance either oscillating around unity, or decreasing as $t^{-n}$ {\red with $n>1$ (or exponentially for $c=1$)} by $t=1000$ 
(see Fig. \ref{fig:MasterPlot2}). The cases which converge slowly (slower than $t^{-0.5}$) as well as those that converge late (after $t=500$) are in the purple zone (the region in the 
center with mostly diamond shaped points for those viewing without colour).

\section{Perturbation Theory near Equilibrium} \label{near}

{\red As we saw above, by construction,} the quantum mechanical equations are approximately recovered, when the spread in phases, specified by $w(a,\phi,t)$ for a given value of the observable $a$, $\Delta\phi_a$, satisfies:
\begin{equation}
\Delta\phi_a \ll \pi, {\rm ~and~} \Delta\phi_a \ll \Delta\phi_F,
\end{equation}
where $\Delta\phi_F$ characterizes the phase spread/width of the kernel $F$. These conditions ensure that different coefficients in the sums: $\sum_\phi w(a,\phi)$  in Eqs. (\ref{eq:phidotnew0}-\ref{eq:rhodotnew0}) factor out, which then let us replace the sums with unity.   In other words, for narrowly-spread phases per observable value,  one can combine Eqs. \ref{eq:phidotnew0}-\ref{eq:rhodotnew0} (or \ref{eq:phidotnew}-\ref{eq:rhodotnew}) for systems, to evolution equations for mean phase and total probability {\it per observable value} to approximately recover quantum mechanics (\ref{eq:origphidot}-\ref{eq:origrhodot}).

In order to examine the next order approximation to compare this evolution to quantum mechanics, we examine the analytical case for a diagonal Hamiltonian. The reason for this choice is that the diagonal case is simple enough to do this calculation analytically, and still can represent any quantum mechanical  system {\red (assuming unitary equivalence, which we discuss in the next section)}. For a diagonal Hamiltonian, Eqs. (\ref{eq:phidotnew0})-(\ref{eq:rhodotnew0}) separates for different observable values, and can be written as:
\bea
\dot{\phi}_i = \sum_j w_j \cos(\phi_i-\phi_j) \sqrt{\sum_k  w_k F(\phi_j-\phi_k)\over  \sum_k w_k F(\phi_i-\phi_k)}, \\
\dot{w}_i = 2w_i \sum_j w_j \sin(\phi_i-\phi_j) \sqrt{\sum_k  w_k F(\phi_j-\phi_k)\times \sum_k w_k F(\phi_i-\phi_k)}, 
\eea
where we have suppressed the dependence on $a$ for brevity, and chosen $R^{-1}_{a a}$ as unit of time. Furthermore, as we saw above, in the regime close to quantum mechanics we can assume $\Delta \phi \ll \pi, \Delta\phi_F$, thus we can Taylor expand both $F$ and cosine:
\bea
\cos(\phi-\phi') \simeq 1-\frac{(\phi-\phi')^2}{2}, \\
F(\Delta \phi) \simeq 1- \frac{\Delta\phi^2}{\Delta\phi_F^2},
\eea
where $\Delta\phi_F^{-2} \equiv -\frac{1}{2} F''(0)$ here. {\red For example, $\Delta \phi_F =2 c^{-1}$ for systems studied in the last section (Eq. \ref{eq:spikeF}).  In this limit, the evolution for $\phi_i$ and $w_i$ can be approximated as:
\bea
\dot{\phi}_{im} &=& \frac{\lambda}{2} \left[(\phi_i - \langle \phi \rangle)^2 -(\phi_m - \langle \phi \rangle)^2\right] + {\cal O}(\Delta \phi^4), \label{dphi_near}\\
\dot{w}_i &=& 2 w_i (\phi_i- \langle \phi \rangle) + {\cal O}(\Delta \phi^3), \label{dw_near}
\eea
where
\bea
\lambda \equiv \Delta\phi^{-2}_F-1, \\
 \phi_{im} \equiv \phi_i-\phi_m,\\
 \langle \phi \rangle \equiv \sum_i w_i\phi_i. \label{phi_ave}
\eea
We now notice that Eq. (\ref{dphi_near}) can be written as:
\beq
\frac{d}{dt}(\phi_i-\langle \phi \rangle) - \frac{\lambda}{2}  (\phi_i-\langle \phi \rangle)^2 = G(t), \label{dphi_near_2}
\eeq 
where $G(t)$ is a yet to be determined function of time. In particular, combining Eqs. (\ref{phi_ave}-\ref{dphi_near_2}) yields:
\beq
G(t) = -\frac{(\lambda+4)}{2} \langle \Delta \phi^2 \rangle =   -\frac{(\lambda+4)}{2}  \sum_i w_i (\phi_i -\langle \phi \rangle)^2. \label{G(t)}
\eeq

The advantage of Eq. (\ref{dphi_near_2}) is that it decouples the evolution for different phases, and thus reduces to a first order ODE for a given $G(t)$. Unfortunately, the solution cannot be written in closed form for arbitrary $G(t)$. 
\subsection{ Power Law Convergence: $\lambda >0$} \label{lambda_pos}

We first introduce an ansatz, which, as we see below, is applicable for $\lambda > 0$, or equivalently $\Delta \phi_F <1$ (implying $c>2$ for the models of Sec. \ref{numeric}).    

To proceed, we next notice a scaling symmetry of the Equations (\ref{dphi_near}-\ref{dw_near}), which remain invariant under:
\beq
\phi_{im} \rightarrow A\times \phi_{im}, ~~t \rightarrow A^{-1} \times t,
\eeq
for arbitrary $A$. Therefore, we postulate an ansatz: $G(t) \propto \langle \Delta \phi^2 \rangle  \propto t^{-2}$. With this assumption, it is convenient to define:
\beq
\tilde{\phi}_i \equiv \lambda t (\phi_i-\langle \phi \rangle), \tau \equiv \ln(t), G(t) = - \frac{\sigma^2-1}{2\lambda t^2}, \label{tilde_def}
\eeq
which let us write  Eq. (\ref{dphi_near_2}) as 
\bea
\tilde{\phi}_i' = \tilde{\phi}_i+\frac{1}{2}\tilde{\phi}_i^2-\frac{\sigma^2-1}{2} &=& \frac{1}{2} (\tilde{\phi}_i+\sigma+1)(\tilde{\phi}_i-\sigma+1). \nonumber\\ &&\label{tilde_phi_dyn}\\
 w_i' &=& 2\lambda^{-1} w_i \tilde{\phi} \label{tilde_w_dyn}
\eea
Note that $\sigma$ is an arbitrary constant, with $\sigma >1 $ for $\lambda >0 $ and $0 < \sigma <1 $ for $\lambda <0 $, which is set by the variance of $\tilde{\phi}$, by combining Eqs. (\ref{tilde_def}) and (\ref{G(t)}):
\beq
\langle \tilde{\phi}^2 \rangle = \frac{\sigma^2-1}{1+4/\lambda} .\label{var_tilde}
\eeq
Moreover, $'$ denotes derivative with respect to $\tau$.  

Interestingly, both the variance of $\tilde{\phi}$ (Eq. \ref{var_tilde}) and its equations of motion (\ref{tilde_phi_dyn}-\ref{tilde_w_dyn}) become time-independent, which potentially admit steady state distributions. In particular, Eq. (\ref{tilde_phi_dyn}) has two fixed points at 
\beq
\tilde{\phi}_{\pm} = -1 \pm \sigma.
\eeq
The fixed point $\tilde{\phi}_+$ is an unstable fixed point, while $\tilde{\phi}_-$ is stable. For positive $\lambda$, it turns out that phases approach infinity (in the perturbative eq's) within a finite time in the interval $(\tilde{\phi}_+,\infty)$, which given the periodic nature of phase, puts them within $(-\infty, \tilde{\phi}_-)$. For negative $\lambda$, the stability and evolution for $\tilde{\phi}$ is identical to that for positive $\lambda$, but this is not the case for $\phi$. Since $\frac{\partial \tilde{\phi}_i'}{\partial \tilde{\phi}_i} = 1 + \tilde{\phi}_i = \pm \sigma$, near the fixed point $\tilde{\phi}_i - \tilde{\phi}_\pm \propto t^{\pm \sigma}$ which is equivalent to $\Delta\phi \propto t^{\pm \sigma-1}+C \tilde{\phi}_\pm t^{-1}$, where $C$ is an unknown constant. Since $\sigma<1$ for negative $\lambda$, this is stable for both fixed points. Therefore, given that we would like to study small phase variances, it stands to reason that we focus on the $(\tilde{\phi}_-, \tilde{\phi}_+)$ interval. Now, defining $w(\tilde{\phi})$ as the weight (or probability) density in the $\tilde{\phi}$ space, we can write the {\it steady-state} continuity equation:
\beq
\cancel{\frac{\partial w}{\partial \tau}}+\frac{\partial }{\partial \tilde{\phi}} \left[ w(\tilde{\phi}) \tilde{\phi}'\right] = w'(\tilde{\phi}),\label{continuity_0}
\eeq
which, plugging from Eqs. (\ref{tilde_phi_dyn}-\ref{tilde_w_dyn}), becomes:
\beq
\frac{\partial }{\partial \tilde{\phi}} \left[ w(\tilde{\phi}) \left( \tilde{\phi}+\frac{1}{2}\tilde{\phi}^2-\frac{\sigma^2-1}{2}\right) \right] = 2\lambda^{-1}  w(\tilde{\phi})\tilde{\phi},\label{continuity}
\eeq 
which can be integrated to give:
\bea
w(\tilde{\phi}) \propto (\tilde{\phi}_+ -\tilde{\phi})^{\alpha_+} (\tilde{\phi} -\tilde{\phi}_-) ^{\alpha_-},\nonumber\\
\alpha_{\pm} = -1+\frac{2}{\lambda} \mp  \frac{2}{\lambda\sigma} .
\eea
We notice a few interesting properties of this solution:
\begin{enumerate}
\item For $\lambda >0$, $\alpha_{\pm} > -1$, which ensures that the probability = $\int w(\tilde{\phi}) d\tilde{\phi}$ is finite. This is not the case for $\lambda <0$, implying that the steady-state solution is non-existent. We can thus consider:
\bea
\lambda = \Delta\phi^{-2}_F-1 = -\frac{1}{2} F''(0) -1 >0, \nonumber \\
\Rightarrow ~-F''(0) > 2
\eea
as a necessary condition to approach equilibrium, at least within the above scaling ansatz:  $G(t) \propto t^{-2}$. 

\item  We further notice that integrating the continuity equation (\ref{continuity}) over our domain $(\tilde{\phi}_-, \tilde{\phi}_+)$,  as the left hand side is a total derivative, and its argument $w(\tilde{\phi}) \tilde{\phi}'$ vanishes at the boundaries (since  $\alpha_{\pm} > -1$). Therefore, the right hand side $\propto \int \tilde{\phi} w(\tilde{\phi}) d\tilde{\phi} = \langle \tilde{\phi} \rangle =0$, which is consistent with our definition of $\tilde{\phi}$ (Eq. \ref{tilde_def}).   

\end{enumerate}

We note that the condition $\lambda >0$ is equivalent to $c>2$, where we see a power law decay of the standard deviation $ \sigma_\phi \propto t^{-1}$ in our numerical simulations (e.g. third row in Table \ref{tab:Plots-of-Equilibrium-2}). Therefore, in spite of the fact that the analysis above is for continuum distributions, it roughly predicts the correct asymptotic behaviour of simple discrete simulations. 

\subsection{Exponential Convergence: $\lambda<0$} \label{lambda_neg}

As we saw above, for $\lambda<0$ (or $\Delta\phi_F >1$),  the power-law ansatz does not lead to a sensible asymptotic steady state distribution.   Let us now propose a different anstaz, i.e. that asymptotically $w(\Delta \phi)$ becomes time-independent, {\it without} additional time re-scaling, where $\Delta \phi \equiv \phi - \langle \phi \rangle $. Moreover, we assume $G(t) = \langle \Delta \phi^2 \rangle =0$.  Again, since there is no explicit time-dependence left in the evolution equations, the continuity equation (\ref{continuity_0}) now reads:
\beq
\cancel{\frac{\partial w(\Delta\phi)}{\partial t}}+\frac{\partial}{\partial \Delta \phi} \left[w(\Delta\phi) \times \frac{\lambda}{2} \Delta \phi^2 \right] = 2 w(\Delta \phi) \Delta \phi,
\eeq
which can be integrated to give:
\beq
w(\Delta \phi) \propto |\Delta \phi|^{4/\lambda-2}.
\eeq
Since the integrals over $w(\Delta \phi) $ are divergent for $\lambda<0$, we have to use a cut-off $\Delta \phi_{\rm min} \rightarrow 0$. After properly normalizing $w(\Delta \phi) $, this yields:
\beq
\langle \Delta \phi^2 \rangle = \left(\frac{\lambda -4}{-\lambda -4} \right) \Delta\phi^2_{\rm min} \rightarrow 0, ~{\rm for}~ -1<\lambda<0 ,
\eeq
consistent with $G(t)=0$ ansatz above. 

Given that this static solution has zero variance, it appears that it cannot have any discrete counterpart. However, one may consider any initial condition $w(\Delta\phi)$ as a perturbation around this static solution, which could be expanded into exponentially decaying modes \footnote{Note that there could be no exponentially growing mode, as the total probability is conserved}. While not entirely rigorous, this analytic argument provides an intuitive understanding of why convergence to quantum mechanics is exponential, rather than power-law, for $\lambda <0$, or $c<2$, e.g. in the second row in Table \ref{tab:Plots-of-Equilibrium-2}. 

\subsection{Approaching Quantum Mechanics}

How does this deviation from quantum mechanics affect physical observable values? Simple manipulations  yield:

\beq
\langle \phi \rangle \dot{}   = \frac{d}{dt} \sum_i w_i\phi_i =  1+ \langle \Delta\phi^2\rangle  +{\cal O} (\Delta\phi^4),
\label{eq:phiavedot}\eeq
More generally, we can write a hierarchy of equations for all the moments of $\Delta \phi$:
\beq
\langle \Delta\phi^m\rangle \dot{} =  (2 + \frac{m\lambda}{2} )\langle \Delta\phi^{m+1}\rangle -\frac{m(\lambda+4)}{2} \langle \Delta\phi^2\rangle\langle \Delta\phi^{m-1}\rangle,
\eeq
which can provide an alternative approach to numerically study convergence to quantum mechanics.	

Instead, here we simply wrap up by making some observations about potential smoking guns of this theory. 
After reintroducing physical units, Eq. \ref{eq:phiavedot} reads:
\beq
\langle \phi \rangle \dot{} - \dot{\phi}_{QM} = \langle \Delta\phi^2\rangle \times \dot{\phi}_{QM},
\eeq
which represents a deviation from quantum mechanics proportional to the variance of phases. This correction can
be absorbed into the Hamiltonian eigenvalues, 
\beq
\tilde{H}_{ii} = H_{ii} \left( 1+\langle \Delta\phi^2\rangle\right), \label{correction}
\eeq
giving an evolving effective Hamiltonian close to the true Hamiltonian. As we saw above (Secs. \ref{lambda_pos}-\ref{lambda_neg}):
\beq
\langle \Delta\phi^2\rangle \propto (Et)^{-2n}, n\geq1. \label{phase_evolve}
\eeq
 Therefore, one expects a time-dependent part of the energy levels of e.g. atoms/nuclei, that decays as $(Et)^{-2}$ or faster asymptotically. }

%

\section{Discussions}\label{discuss}

Many of the issues with the original real ensemble model, presented in  Section VI of \cite{smolin2011real}, still remain open. The only exception is 
the stability of the quantum mechanical equilibrium, which we have demonstrated for a certain class of the non-equilibrium
equations. Here, we discuss some of the other conceptual challenges (or features) in Smolin's real ensemble extension of the quantum theory:

\begin{figure}[H]
\includegraphics[width=0.70\paperwidth]{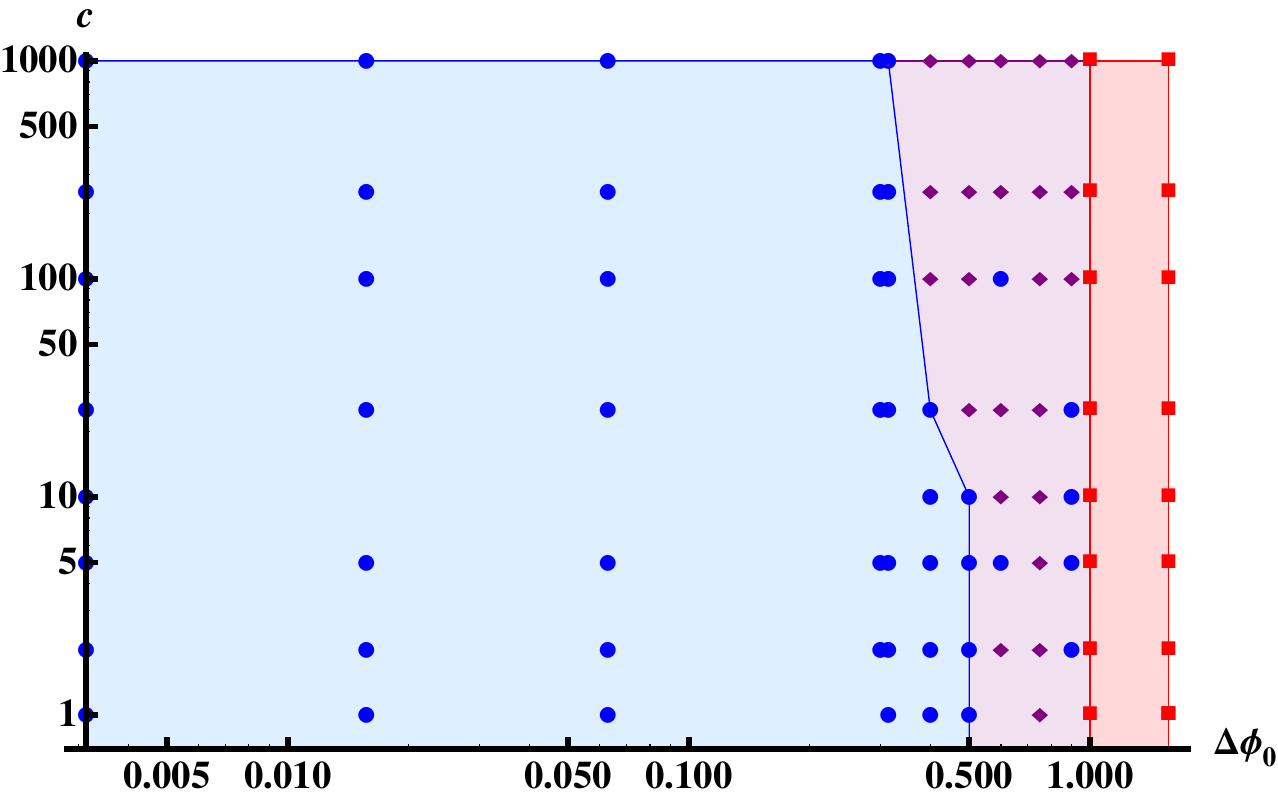}

\caption{\label{fig:MasterPlot}Phase space of convergence to quantum mechanics for $H=2\sigma_{z}$, $\rho\left(0\right)=\left\{ \left\{ 0.16,0.08,0.06\right\} ,\left\{ 0.23,0.3,0.17\right\} \right\}$, $\phi\left(0\right)=\left\{ \left\{ 0,\Delta\phi_{0},2\Delta\phi_{0}\right\} ,\left\{ \frac{\pi}{2}+\Delta\phi_{0},\frac{\pi}{2},\frac{\pi}{2}+\frac{\Delta\phi_{0}}{2}\right\} \right\}$, and $F$ given by Equation \ref{eq:spikeF} with varying $c$ {\red (characterizing the compactness of the F-kernel)} and $\Delta\phi_{0}$ {\red (the initial phase difference per value of the observable)}. Blue circles indicates convergence, red squares indicates lack of convergence, and purple diamonds means spin up converges, but spin down does not.}

\end{figure}

\subsection{Zero Occupancy States}

An interesting aspect of this model is the fact that if $\rho_{a}=0$,
$\dot{\rho}_{a}=0$ and this potential observable value is effectively non-existent. This
is necessary as Equation \ref{eq:origphidot} breaks down when $\rho_{a}=0$. In
the non-equilibrium extension, this becomes $\rho_i=0$
meaning $\dot{\rho}_i=0$. Despite the 
$\underset{k}{\sum}\rho_k \delta_{a_i a_k}$
term in the denominator of Equation \ref{eq:rhodotnew}, it can be seen that
this remains true even if all $\rho_i$ terms
are zero for a given value of $a_i$. This will effectively eliminate
a set of beables from being possible if at any time in its evolution,
that set of beables has a zero probability. This ensures the quantum
mechanical case is a fixed point, as if each value of the observable only has one
potential phase, no other potential phases will be able to become
non-zero, and the condition of one phase per value of the observable is maintained.
This also causes a non-quantum mechanical issue at the nodes of a
quantum mechanical system in the cases where in quantum mechanics
a node ceases to be a node at later times due to the evolution. 

This model is also by nature discrete. This implies two things for
the property above. First is the finiteness of the number of sets
of beables, which will cause issues with the examination of continuous
and infinite systems as will be examined in the next section. The
second is what happens in a large $N$ system for pairs of beables or times
with small enough probability such that $N$ is no longer large enough
to prevent the discreteness from effecting the system and its behaviour.
This is not just at a node of exactly zero probability, but near one
of very small probability. In this case finite behaviours become important,
and there is a potential for {\red complete loss of} a beable pair by chance,
even if the continuous probability dictates that the probability is
never exactly zero. 

\begin{table}
\caption{\label{tab:Plots-of-Equilibrium-2}Plots of the standard deviation of the phases for the cases in Figure \ref{tab:Plots-of-Equilibrium}.
From top to bottom, the functions $F$ within equations \ref{eq:phidotnew}
and \ref{eq:rhodotnew} for the plots are $F=1$, $F=\cos^{2}\left(\frac{\Delta\phi}{2}\right)$,
and $F$ given by equation \ref{eq:spikeF} with $c=100$. The first two plots have a logarithmic
scaling for the y-axis, while the scaling of both axes are logarithmic for the third plot.}

\begin{tabular}{|c|c|}
\hline 
 & Standard Deviation of Phase vs Time\tabularnewline
\hline
\hline 
\includegraphics[width=0.0346\paperwidth]{ALzz2} & \includegraphics[width=0.60\paperwidth]{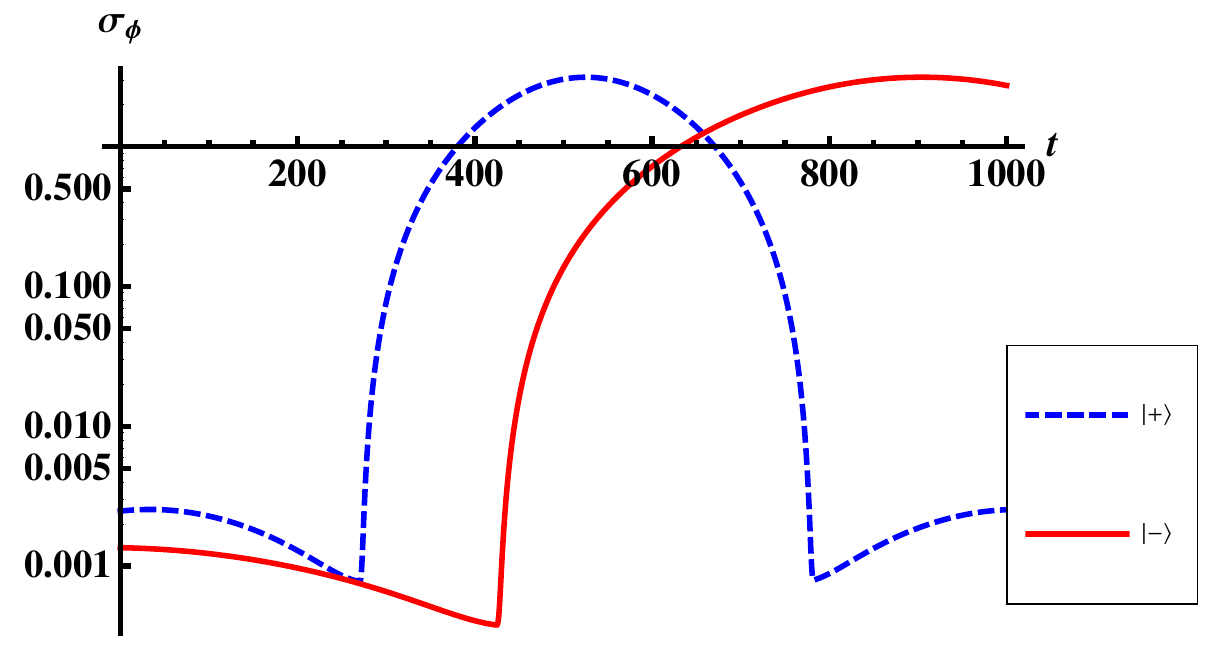}\tabularnewline
\hline 
\includegraphics[width=0.0346\paperwidth]{ALzz2} & \includegraphics[width=0.60\paperwidth]{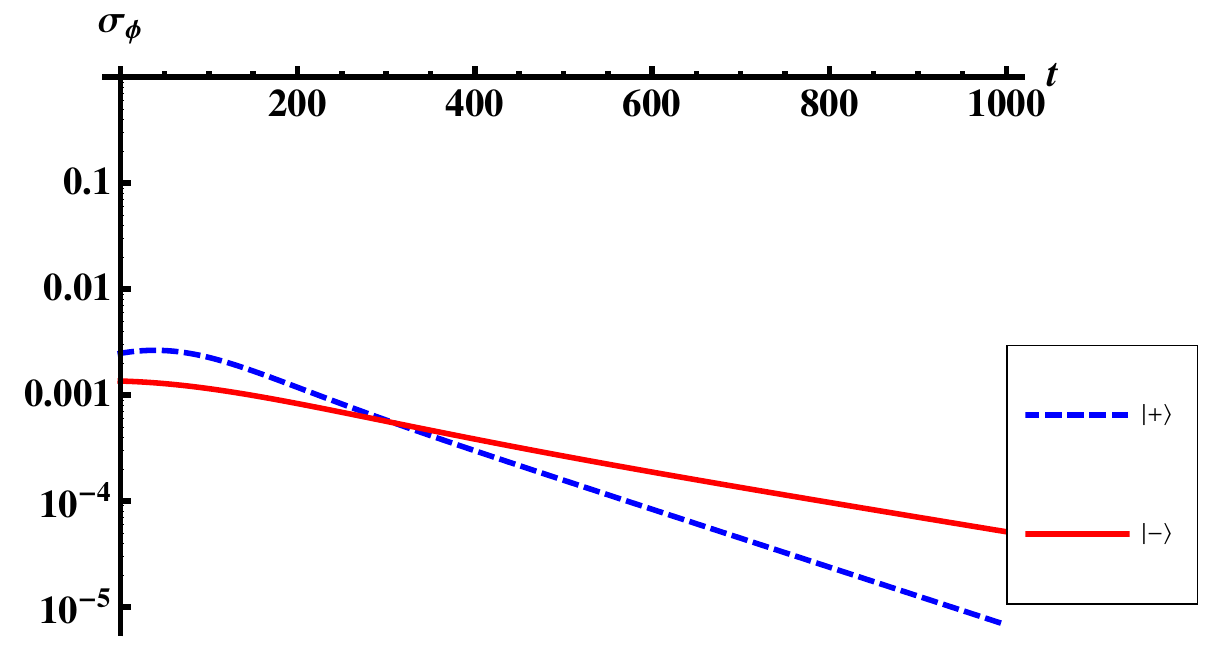}\tabularnewline
\hline 
\includegraphics[width=0.0346\paperwidth]{ALzz2} & \includegraphics[width=0.60\paperwidth]{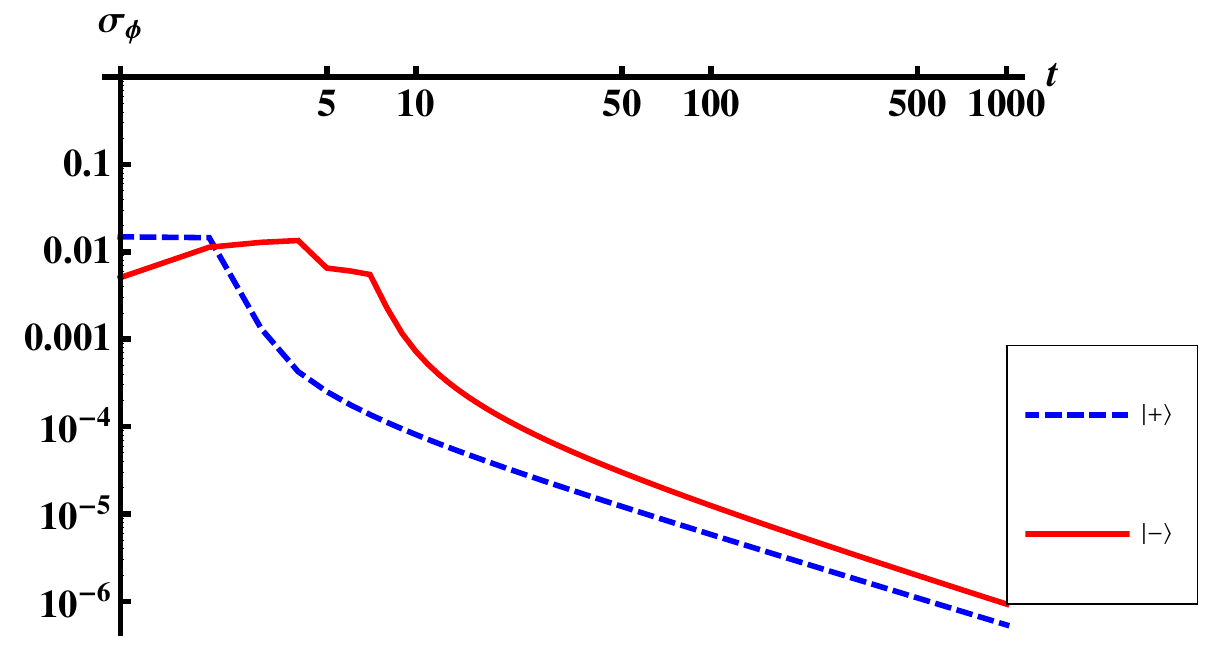}\tabularnewline
\hline
\end{tabular}
\end{table}

\subsection{Unitary Equivalence vs Preferred Basis}\label{unitarity}

Technically, our current model chooses a preferred {\red basis of the Hilbert space, defined by the eigenstates of the observable that is realized in  the real ensemble theory}. However, this would imply {\red that, e.g. only spin in one direction can be measured - it does not account/allow for choice
of 
non-commuting observables, which is part of what contrasts quantum and
classical mechanics.} Since we would like to account for the existence
of non-commuting observables, we need to ask what the model would
give if one were to measure a different observable. To do this, the
model would either have to define a preferred basis for the theory
from which other measurements could be determined or one has to have
a definite observable value for many bases. Both proposals have flaws.
Since this model is discrete by nature (only continuous by approximation),
taking position, or even momentum in a continuous spacetime would
{\red suffer from } the ``node''
problem mentioned above- the probabilities would be spread too thin, 
{\red exposing} the discreteness of the model, causing certain observable values
to be unrepresented, and therefore never represented. In order to
take one of these as the preferred basis as might be desired, a discrete
space-time is required. 

Taking space-time discrete and position space, momentum space, Hamiltonian,
or all three together as the preferred basis/bases, might help if N
is large enough. Any discrete eigen-space would work - e.g. that of the Hamiltonian. A discrete but infinite system would not be able to have
all eigenstates represented though, {\red for a finite N}. 

Even for a finite system, however, if we choose to maintain unitarity,
the fact that the system is finite is again a problem. There is an
infinite number of unitary transformations, unless somehow we discretize
this, for instance by having discreteness in angles for the case of
the spin-1/2 system. So we must take a finite set of preferred bases,
something which in this case, unlike the infinite case, does create
a finite number of types of members of the ensemble in the equilibrium model. 

In continuing with this, a finite number of represented phases per
value of each observable is also required to maintain the fact the system is finite.

The most obvious choice is  to take the Hamiltonian eigenspace 
as preferred, as it exists in all cases and is  native to the equations.
Of course, this doesn't make it the correct, but rather the simplest possible choice. It
is also discrete for finite systems. 
An issue remains for degeneracies in the
eigenstates of this basis, where the Hamiltonian fails to single out a basis, unless the degeneracies are only an idealization/approximation
of the system. This choice has both the advantage and disadvantage
of simplifying the equations. The advantage being that the simplified
system is easy to work with and understand. The disadvantage being that
no equilibrium evolution is seen and parts of the equations remain
unused - parts that match quantum mechanics in equilibrium and enabled
the conclusion that this set of equations matches quantum mechanics. 

\subsection{Real Ensemble Model and Early Universe}

{\red 
Given the experimental success of quantum mechanics in explaining microscopic phenomena, any phenomenologically viable real ensemble theory must be  stable to perturbations away from quantum mechanics, at least within the tested regimes. Indeed, as we have demonstrated in this paper, this is possible for a  large space of parameters in our formulation. Therefore, any smoking gun for this theory should be sought beyond the current empirical regime, i.e. very high (or possibly very low) energies, or prior to convergence to quantum mechanics. Both these conditions are met in the early universe, which implies that one could potentially search for signatures of deviations from quantum equilibrium in cosmological observations. 

As we pointed out in the introduction, indeed one of the motivations for studying hidden variable theories away from quantum equilibrium is non-local signalling that could provide a resolution to the cosmological horizon problem \cite{valentini2002signal}, as well as observed super-horizon correlations in the CMB. Let us see how this could happen in our framework:

The spectrum curvature fluctuations in the radiation era is given by (e.g., see \cite{Geshnizjani:2011dk}):
\beq
P_{\zeta}(k) = \frac{(1+2 n_k) }{(4\pi)^2\sqrt{3}} \left(k\over M_p\right)^2, \label{power}
\eeq
where $n_k$ is the mean occupation number of the mode $k$, and $M_p$ is the Planck mass. In the ground state of quantum mechanics $n_k=0$. However, as we saw in Eq. (\ref{correction}) the phase variance introduces a correction to the effective energy of all the energy eigenstates, including the ground state.  Given that energy of a harmonic oscillator scales as $n_k+\frac{1}{2}$, we may interpret this correction as:
\beq
n_k = \frac{1}{2} \langle \Delta\phi^2 \rangle \simeq \frac{ \langle \Delta\phi^2 \rangle_0 }{ 2 (E_k t)^2}, 
\eeq
where we used the results of Sec. \ref{near} (e.g., Eq. \ref{phase_evolve}) for the asymptotic scaling of the phase variance. Furthermore, for the ground state: 
\beq
E_k = \frac{1}{2} \omega_k = \frac{c_s k}{2} = \frac{k}{2\sqrt{3}},
\eeq
where we used $c^2_s=1/3$ for the radiation fluid. Plugging this into Eq. (\ref{power}) yields:
\beq
P_{\zeta}(k) = \frac{(k/M_p)^2}{16\pi^2 \sqrt{3}}+ \frac{\sqrt{3} \langle \Delta\phi^2 \rangle_0}{8\pi^2 (M_p t)^2}. \label{power_fit}
\eeq
We thus see that the correction term to the $\zeta$ power spectrum is indeed scale-invariant, and assuming  $t \sim  10^{4} \sqrt{ \langle \Delta\phi^2 \rangle_0} M^{-1}_p$, yields $P_\zeta \sim 10^{-9}$, which is roughly consistent with  cosmological observations.  

Of course, the interpretation of the result in Eq. (\ref{power_fit}) is far from straightforward. For example, $P_\zeta$ should become independent of time on superhorizon scales in standard cosmology. So, at what time, $t$,  should we expect the evolution of $\zeta$ to freeze, if at all? This will be a necessary condition for the scale-invariant term to not be negligible at late times. 

A more serious problem with interpreting Eq. (\ref{power_fit}) as the origin of cosmological primordial power spectrum, is that it requires $ \langle \Delta\phi^2 \rangle \gg 1$ to match the amplitude of CMB anisotropies, thus invalidating the near-equilibrium approximation. Therefore, we should only consider this result as a suggestive direction for further exploration of the implications of the real ensemble theory for early universe, pending a deeper understanding of its physics. 

\subsection{Real Ensemble Model, Quantum Gravity, and Cosmological Constant Problem}

A surprising feature of the real ensemble theory that was discovered here, is that the rate of convergence to quantum mechanics depends on the absolute (and not relative) energy. This suggests a possible connection to gravitational physics, as gravity is the only interaction that is sensitive to the total energy of a system. In other words, total energy of a system sources both its gravity, AND its convergence to quantum mechanics.   

As an example, let us consider the expectation for the vacuum energy, or the cosmological constant. In Eq. (\ref{correction}), we saw that the effective energies of eigenstates receive corrections as:
\beq
E = E_{QM} + \frac{ \langle \Delta\phi^2 \rangle_0 }{ E_{QM} t^2},
\eeq
if we interpret $\langle \dot{\phi} \rangle$ as energy, $E$. Now, this energy has a minimum:
\beq
E_{\rm min} \sim \frac{ 2\sqrt{\langle \Delta\phi^2 \rangle_0} }{t}.
\eeq
Now, if we interpret this as vacuum energy within a Hubble patch of the universe:
\beq
\rho_{\rm vac} \sim E_{\rm min} H^{3} \sim (14 ~{\rm meV}^4) \sqrt{\langle \Delta\phi^2 \rangle_0}~T({\rm TeV})^8,
\eeq
which is comparable to the observed dark energy density, if we set $t$ to the time of electroweak  phase transition, or temperature $T \sim$ TeV  (and $\langle \Delta\phi^2 \rangle_0 \sim 1$):

  This might be a reasonable scale to plug in here, as quantum field theory is not well probed beyond TeV scale.  

}

\subsection{Alternative Non-Equilibrium Real Ensemble Model}

In converting from equations \ref{eq:phidotnew0} and \ref{eq:rhodotnew0} to equations \ref{eq:phidotnew} and \ref{eq:rhodotnew}, 
we note an additional possible extension of the model outside of the quantum mechanical case. This case comes from factoring out
a $\rho_a$ term from the square root to remove this term from the denominator of the weight function before taking the equation out
of equilibrium. It is equivalent to the case when all instances of $\underset{k}{\sum}\rho_k \delta_{a_i a_k}$ are replaced with 
$\tilde{\rho}_{a_i}$. The resulting equations governing the model are:
\begin{equation}
\dot{\phi}_i = \underset{j}{\sum}
\frac{\rho_j}{\sqrt{\tilde{\rho}_{a_j}\tilde{\rho}_{a_i}}}
\times R_{a_{i} a_{j}}\cos\left[\phi_{i}-\phi_{j}+\beta_{a_{i} a_{j}}\right],
\end{equation}
\begin{equation}
\dot{\rho}_i = 
\underset{j}{\sum}
\frac{\rho_i \rho_j}{\sqrt{\tilde{\rho}_{a_j}\tilde{\rho}_{a_i}}}
\times2R_{a_{i} a_{j}}\sin\left[\phi_{i}-\phi_{j}+\beta_{a_{i} a_{j}}\right].
\end{equation}

Several aspects make this model appear simpler. The square root denominator is the same for both $\dot{\phi}$ and $\dot{\rho}$. There is only one type
of $\tilde{\rho}$ term. In addition, the weighting based on $\rho_i$ is obvious and separate from the square root terms. This change comes from acknowledging 
this implicit weighting from the equilibrium model, and separating it from the square roots.

Numerically, the new model gives results close to, if not effectively identical to, the original model when analyzing for convergence.
This means that although the exact evolution may differ, all result obtained for the original non-equilibrium model remain valid. The convergence is invariant to
this change in extrapolation.

Both models are identical when $F=1$.

\section{Conclusions}\label{conclude}

{\red Inspired by theoretical and observational motivations for violating relativistic locality,  we explored the non-local real ensemble model, in the context proposed by Smolin \cite{smolin2011real}. In particular, the theory generalizes quantum mechanics by allowing a range of (or uncertainty in) quantum phase per value of an observable. 

We first developed the generalized evolution for the real ensemble, which consist of continuous phase evolution, and copy rules.  We then provided both numerical and analytic evidence for why quantum mechanics (with one phase per observable value) is a local stable fixed point for most of the parameter space of the theory. Moreover, the phase variance, as well as deviations from quantum mechanical frequencies decay faster than $(Et)^{-2n}$, with $n\geq 1$, for stable models. The energy scale of this decay is the absolute energy rather than relative energy of the system, which suggests a possible connection to gravitational physics.

Finally, we discussed different conceptual aspects of the real ensemble theory. This includes the need for a preferred basis  in the Hilbert space (and/or further interpretation), as well as 
potential novel applications for the spectrum of primordial cosmological perturbations and the cosmological constant problem.
}
%
%
%

\begin{acknowledgements}
 We would like to thank Lee Smolin for inspiring this work, and many discussions along the way. We also thank Lucien Hardy, and Steve Weinstein for comments on the draft. This work was supported by the Natural Science and Engineering Research Council of Canada,
the University of Waterloo and by Perimeter Institute for Theoretical Physics. Research
at Perimeter Institute is supported by the Government of Canada through Industry
Canada and by the Province of Ontario through the Ministry of Research \& Innovation. 
\end{acknowledgements}

\appendix
\section{Various Plots of Numerical Simulations}
Here are given additional plots of the numerical simulations for those that wish to confirm certain aspects covered in the paper. 

\begin{table*}
\caption{\label{tab:Plots-of-Equilibrium-1-1}Plots of the evolution of spin-$\frac{1}{2}$
systems in the non-equilibrium real ensemble model for the case with a moderate initial separation of phases. Each case has
three different phases for each of the two potential values of $s_z$. The initial
conditions are $\rho\left(0\right)=\left\{ \left\{ 0.16,0.08,0.06\right\} ,\left\{ 0.23,0.3,0.17\right\} \right\} $
if marked as uneven or $\rho\left(0\right)=\left\{ \left\{ 0.2,0.1,0.2\right\} ,\left\{ 0.2,0.1,0.2\right\} \right\} $
if marked as even and $\phi\left(0\right)=\left\{ \left\{ 0,0.1\pi,0.2\pi\right\} ,\left\{ \frac{\pi}{2}+0.1\pi,\frac{\pi}{2},\frac{\pi}{2}+0.05\pi\right\} \right\} $.
The Hamiltonian is $H=\sigma_{x}+\sigma_{z}$ for the first two plots and $H=2\sigma_{z}$ for the third. The functions $F$ within equations \ref{eq:phidotnew}
and \ref{eq:rhodotnew} for the plots are $F=\cos^{2}\left(\frac{\Delta\phi}{2}\right)$
for the top plot and $F$ given by equation \ref{eq:spikeF} with
$c=100$ for the other two.}

\begin{tabular}{|c|c|c|c|}
\hline 
 & Probability vs Time & Total Probability vs Time & Phase Difference vs Time\tabularnewline
\hline
\hline 
\includegraphics[width=0.015\paperwidth]{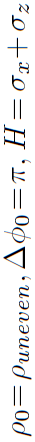} & \includegraphics[width=0.26\paperwidth]{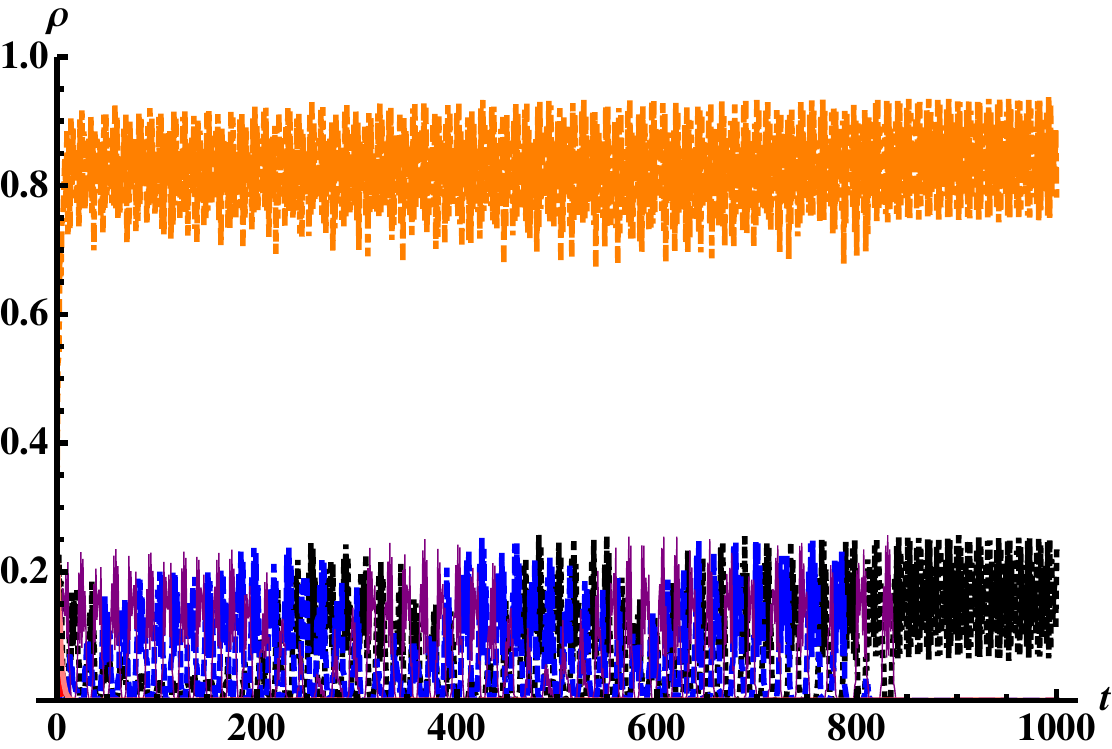} & \includegraphics[width=0.26\paperwidth]{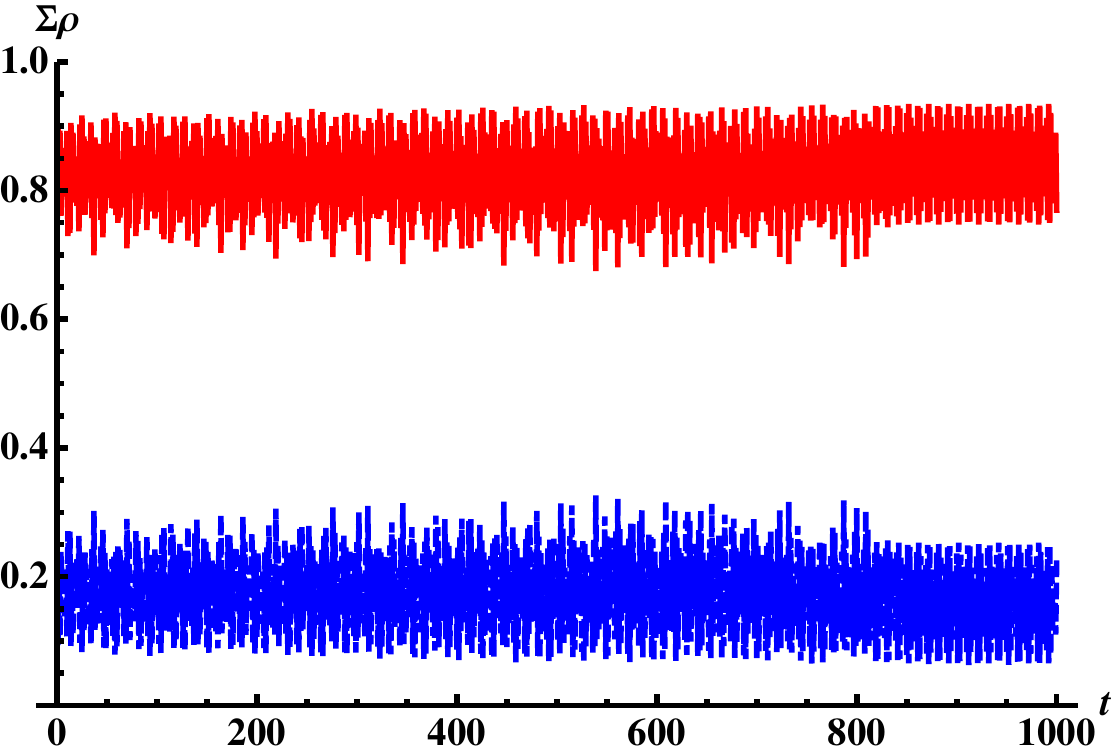} & \includegraphics[width=0.26\paperwidth]{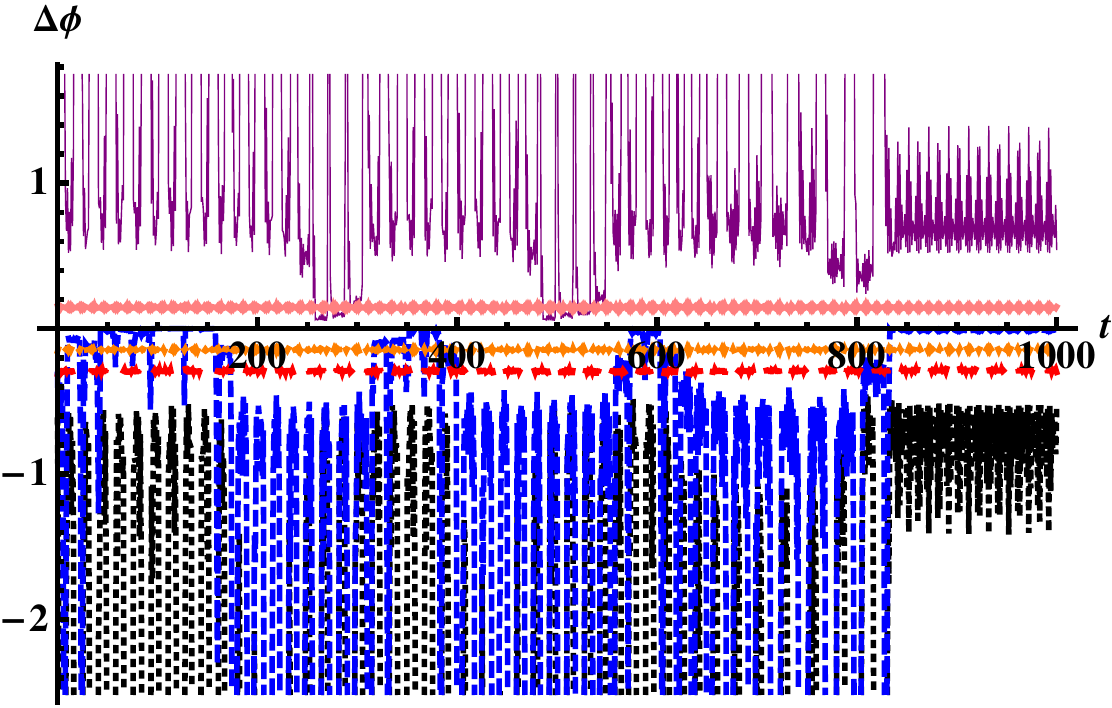}\tabularnewline
\hline 
\includegraphics[width=0.015\paperwidth]{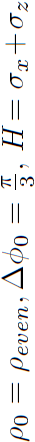} & \includegraphics[width=0.26\paperwidth]{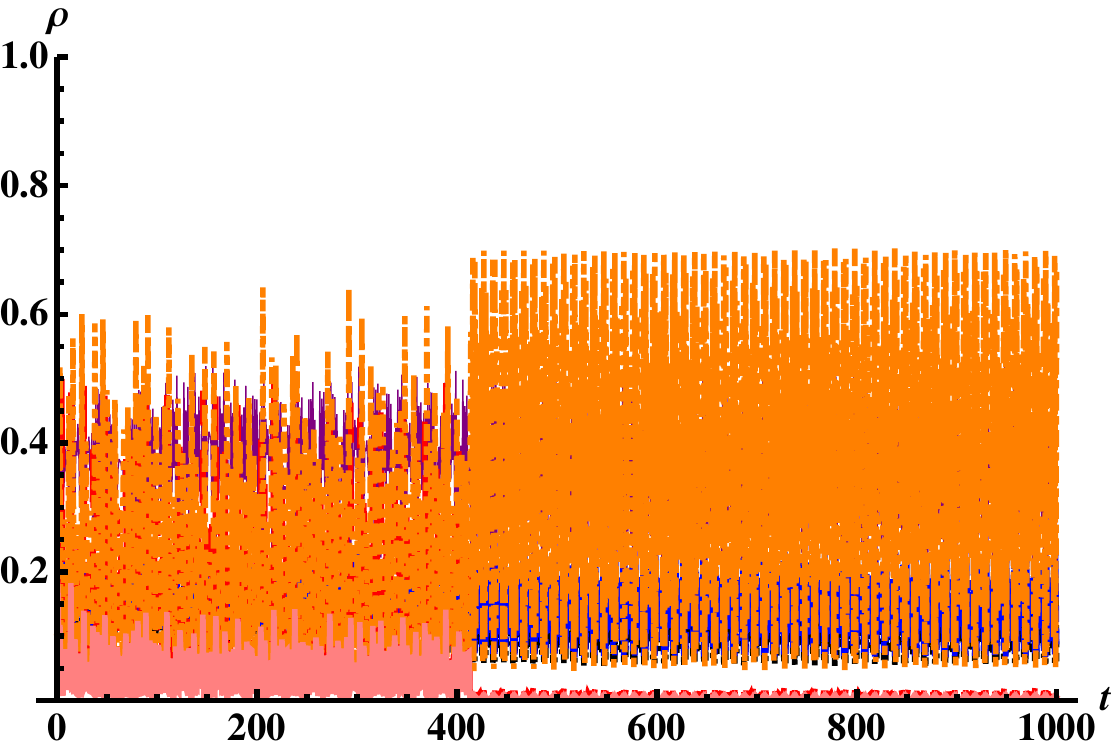} & \includegraphics[width=0.26\paperwidth]{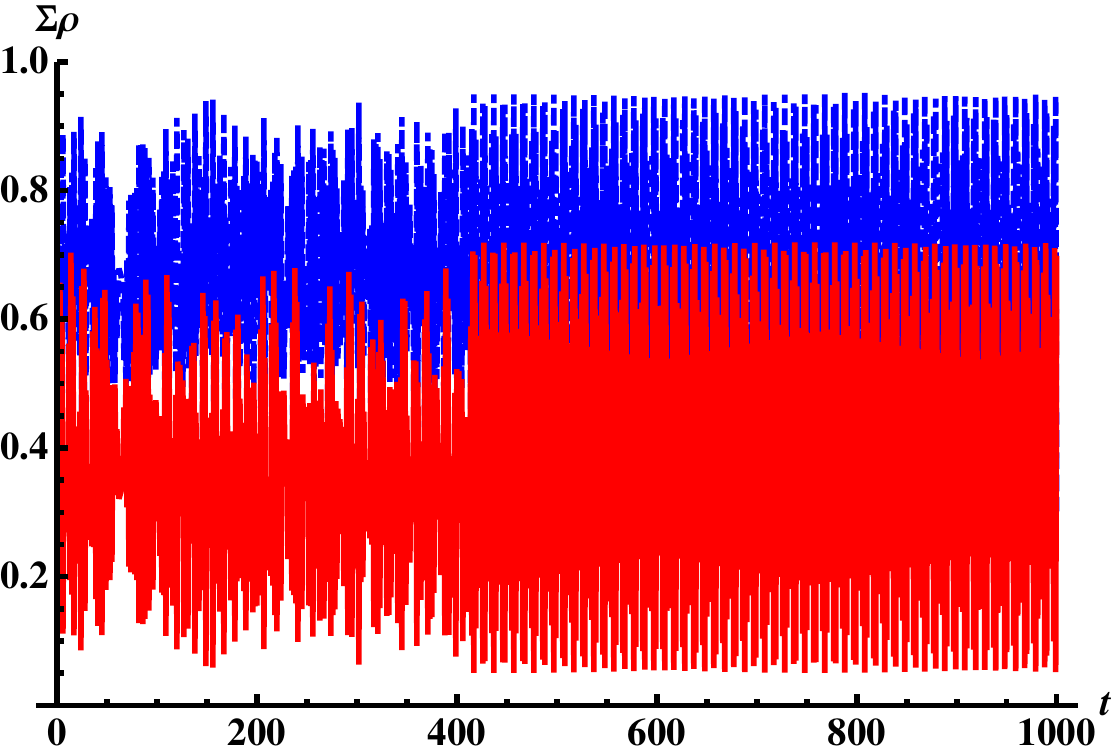} & \includegraphics[width=0.26\paperwidth]{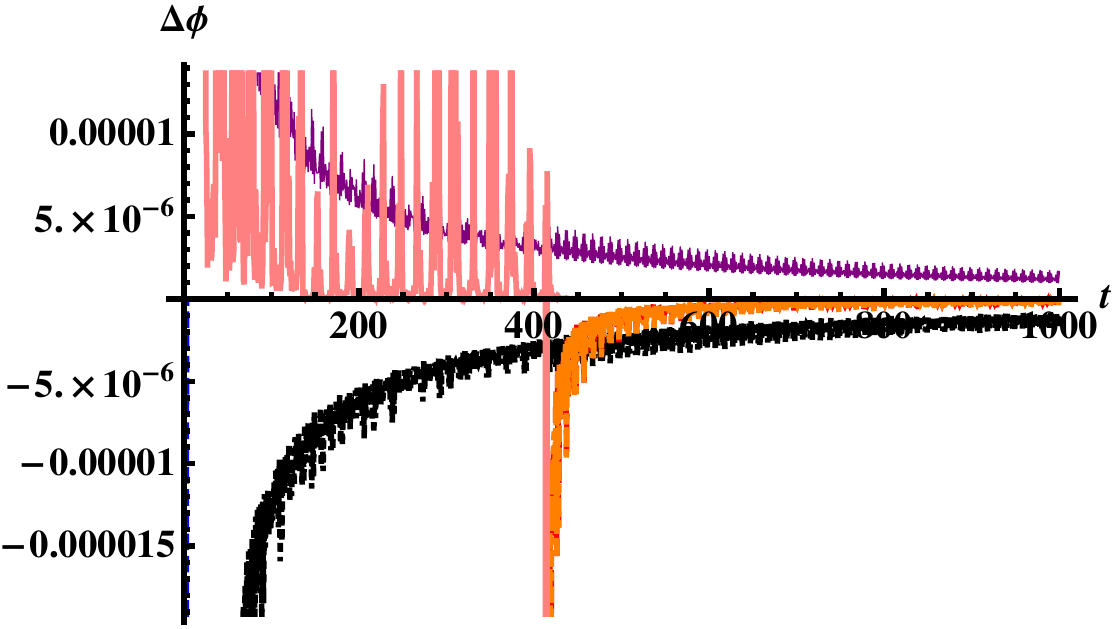}\tabularnewline
\hline 
\includegraphics[width=0.015\paperwidth]{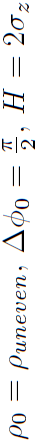} & \includegraphics[width=0.26\paperwidth]{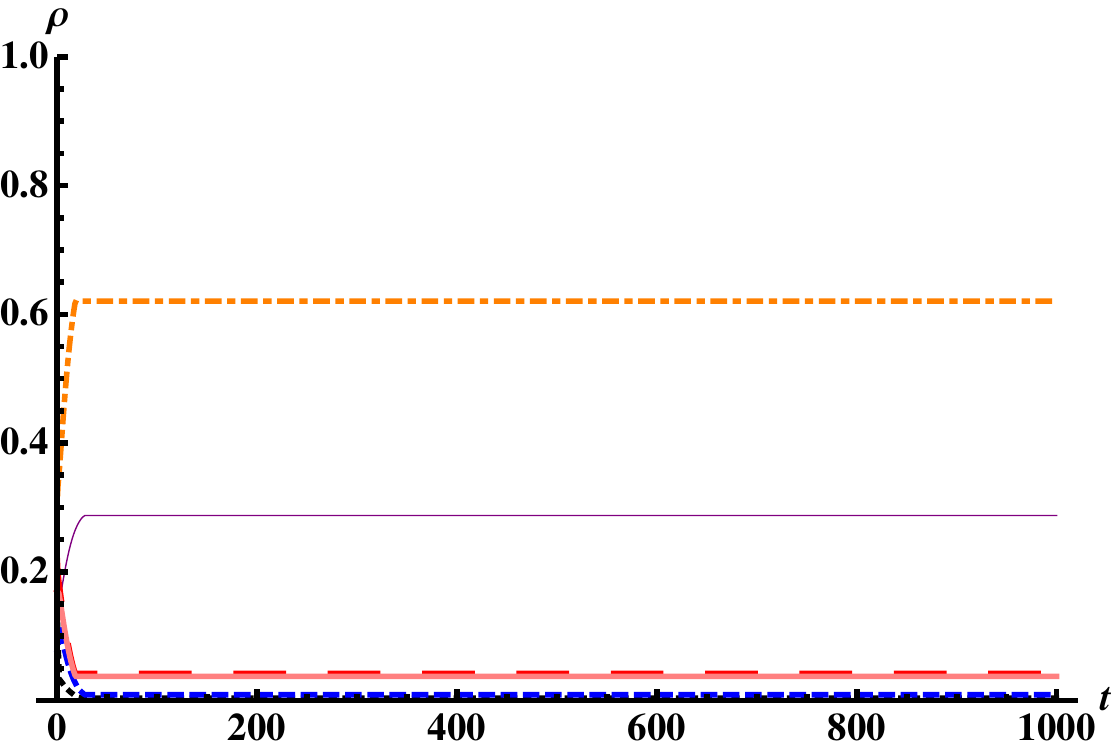} & \includegraphics[width=0.26\paperwidth]{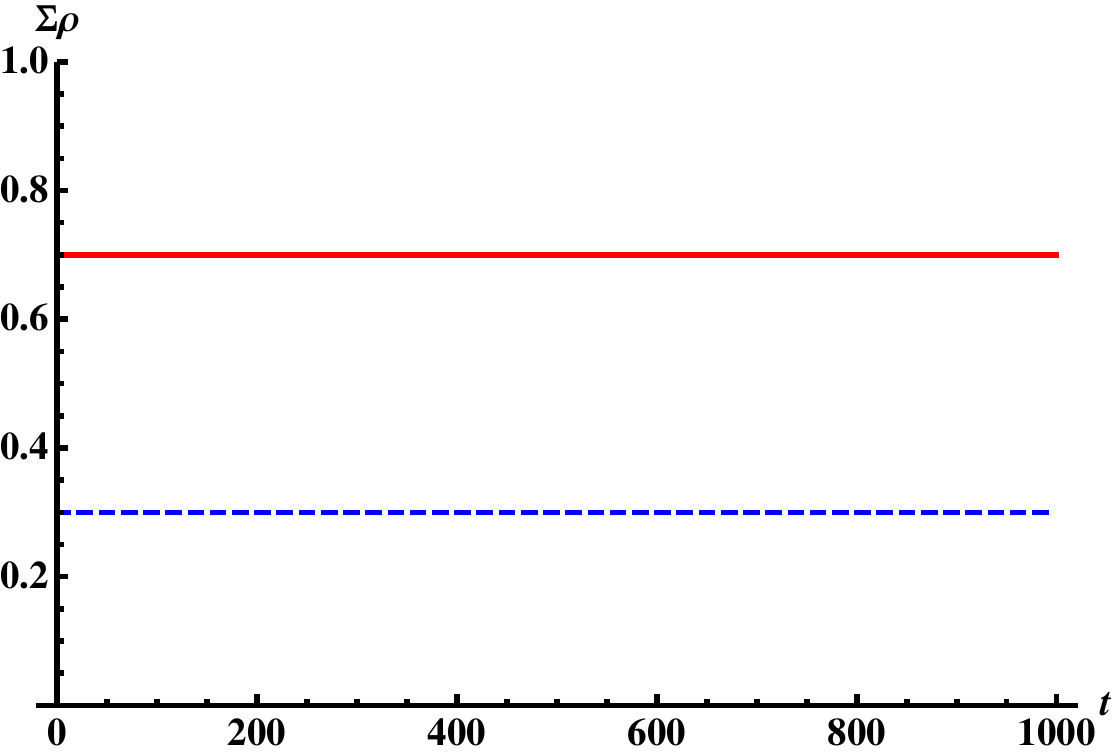} & \includegraphics[width=0.26\paperwidth]{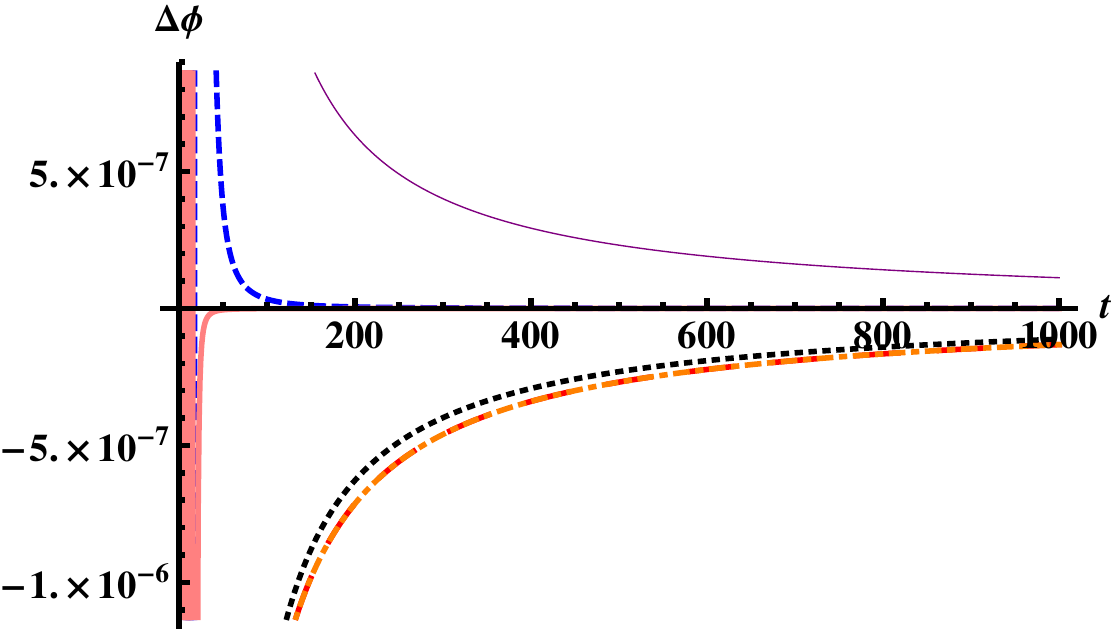}\tabularnewline
\hline
\end{tabular}
\end{table*}

\begin{table*}
\caption{\label{tab:Plots-of-Equilibrium-1}Plots of the evolution of spin-$\frac{1}{2}$
systems in the non-equilibrium real ensemble model for the case of large initial separation of phases. Each case has
three different phases for each of the two potential values of $s_z$. The initial
conditions are $\rho\left(0\right)=\left\{ \left\{ 0.16,0.08,0.06\right\} ,\left\{ 0.23,0.3,0.17\right\} \right\} $
and $\phi\left(0\right)=\left\{ \left\{ 0,1,2\right\} ,\left\{ \frac{\pi}{2}+1,\frac{\pi}{2},\frac{\pi}{2}+0.5\right\} \right\} $.
The Hamiltonian is $H=\sigma_{x}+\sigma_{z}$ for the top plot and $H=2\sigma_{z}$ for the bottom plot. From top to bottom, the functions $F$ within equations
\ref{eq:phidotnew} and \ref{eq:rhodotnew} for the plots are $F=\cos^{2}\left(\frac{\Delta\phi}{2}\right)$
and $F$ given by equation \ref{eq:spikeF} with $c=100$. }

\begin{tabular}{|c|c|c|c|}
\hline 
 & Probability vs Time & Total Probability vs Time & Phase Difference vs Time\tabularnewline
\hline
\hline 
\includegraphics[width=0.015\paperwidth]{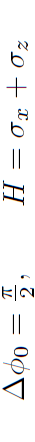} & \includegraphics[width=0.26\paperwidth]{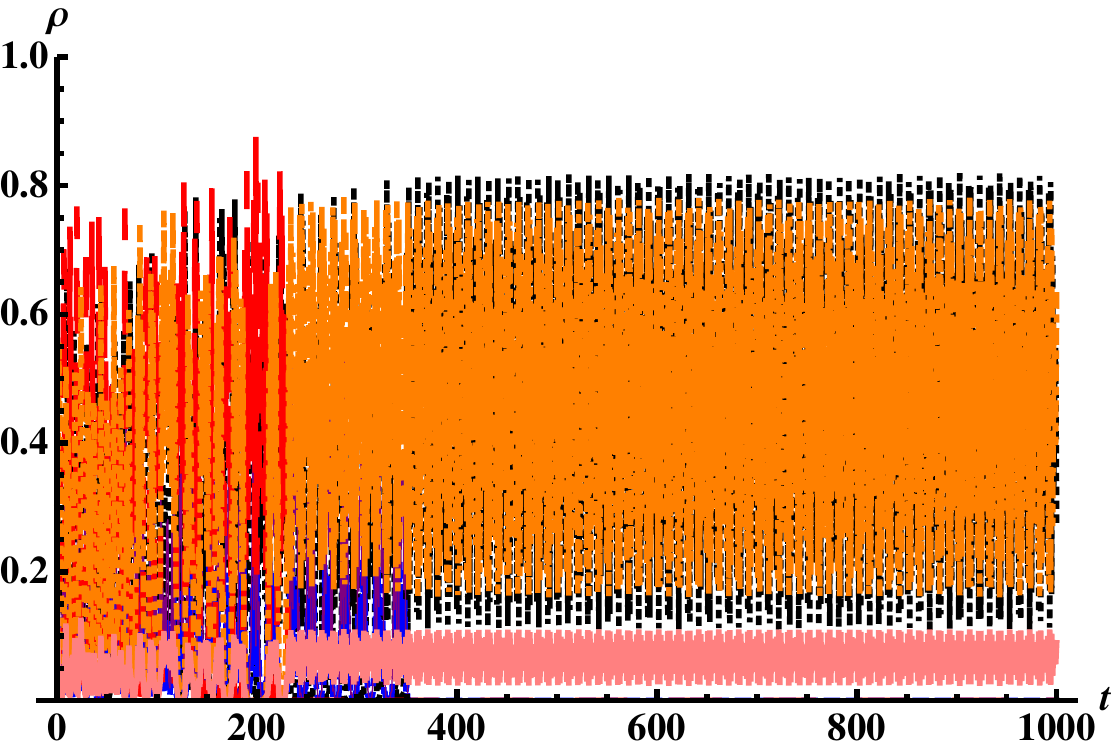} & \includegraphics[width=0.26\paperwidth]{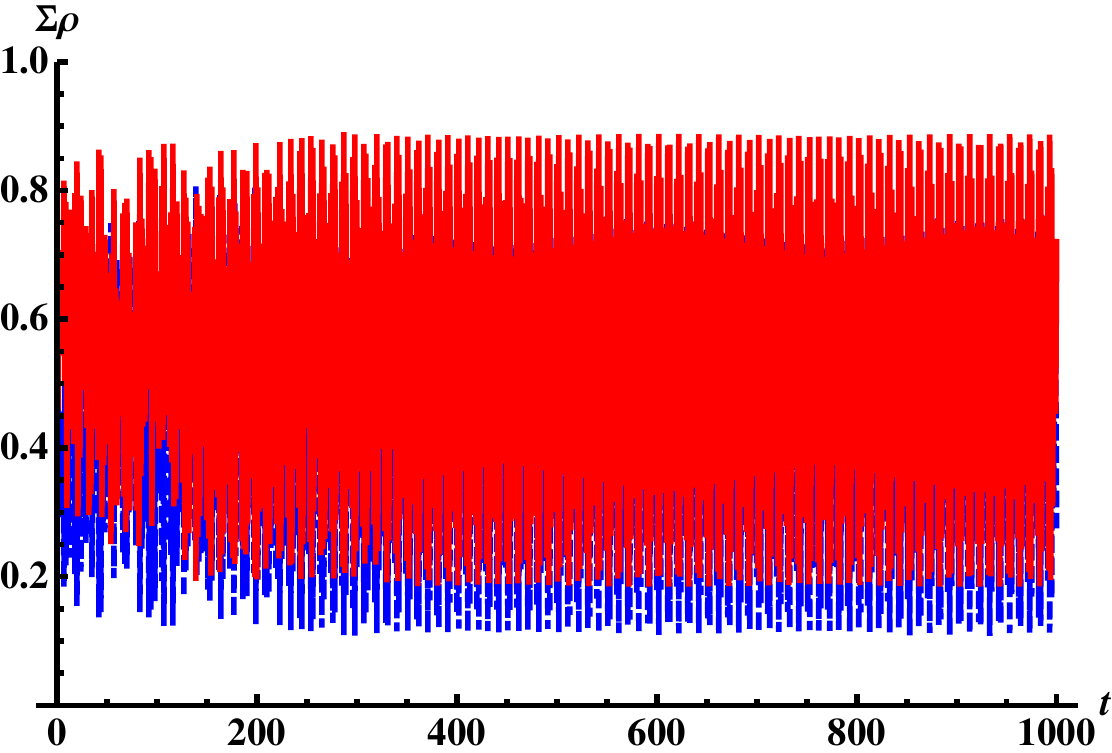} & \includegraphics[width=0.26\paperwidth]{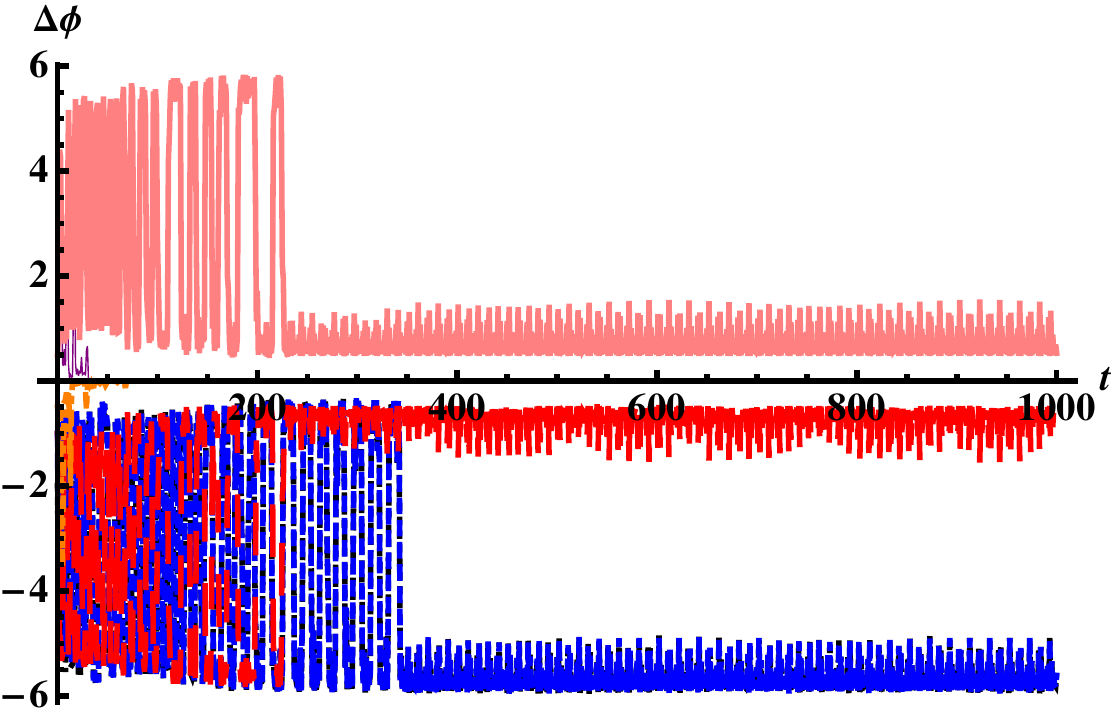}\tabularnewline
\hline 
\includegraphics[width=0.0173\paperwidth]{ALzz2} & \includegraphics[width=0.26\paperwidth]{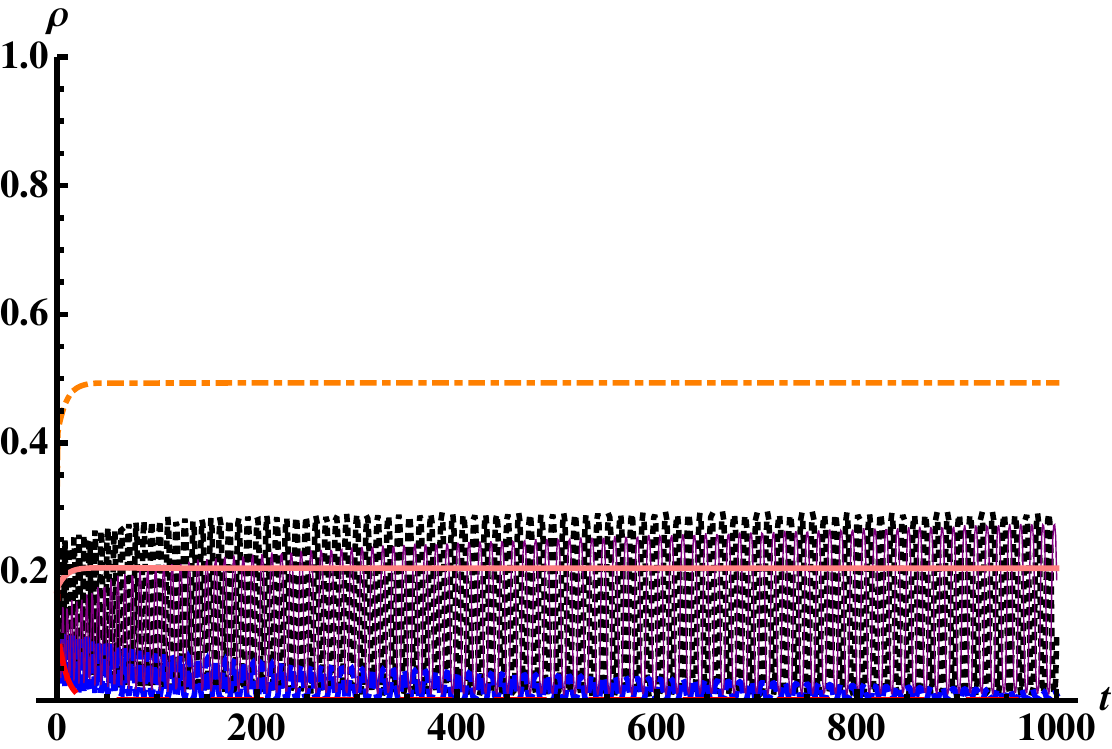} & \includegraphics[width=0.26\paperwidth]{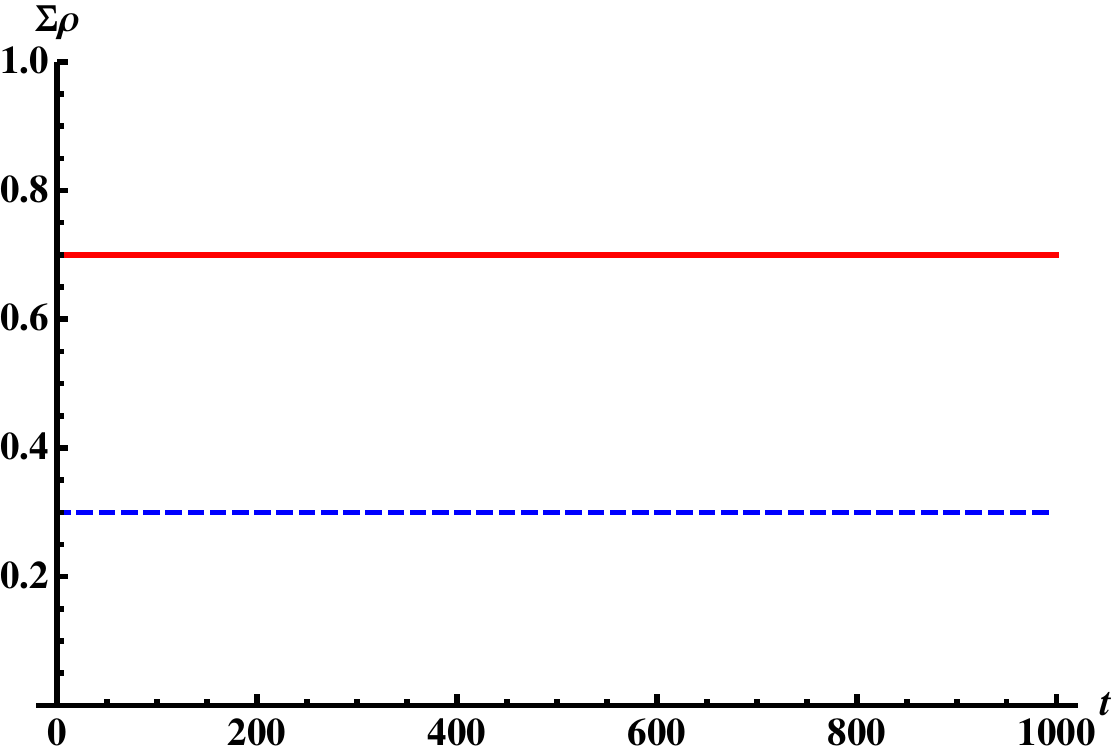} & \includegraphics[width=0.26\paperwidth]{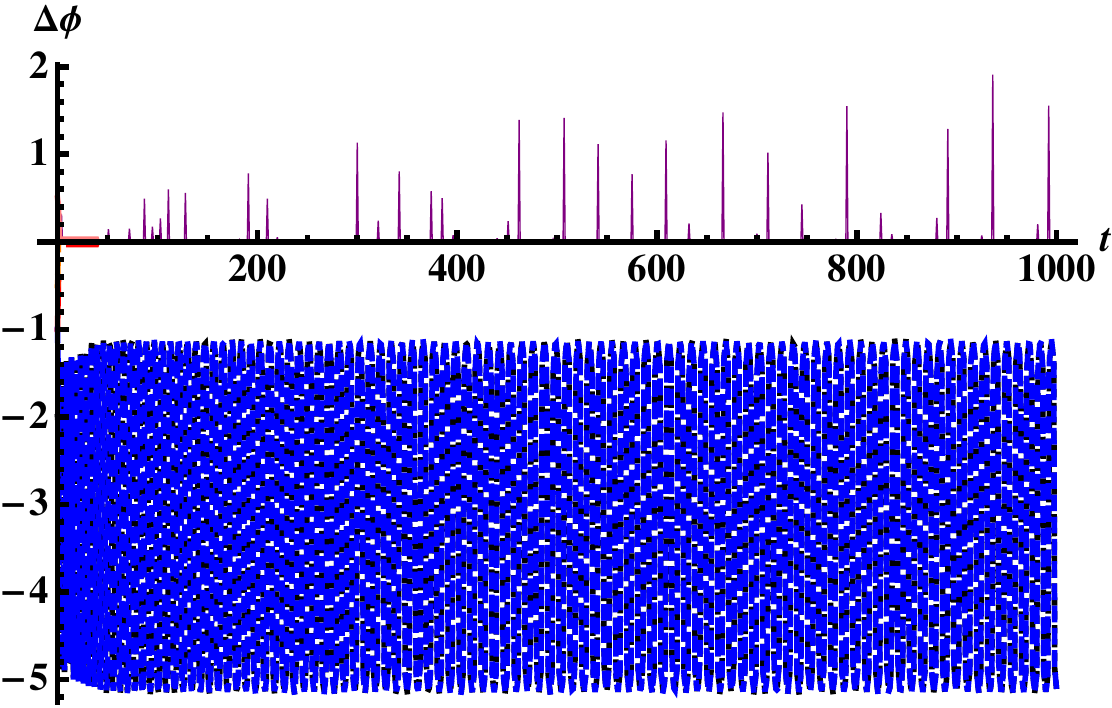}\tabularnewline
\hline
\end{tabular}
\end{table*}

\begin{table*}
\caption{\label{tab:Plots-of-II}Plots of the evolution of spin-$\frac{1}{2}$
systems in the non-equilibrium real ensemble model for the case with a Hamiltonian proportional to the identity. Each case has
three different phases for each of the two potential values of $s_z$. The initial
conditions are $\rho\left(0\right)=\left\{ \left\{ 0.16,0.08,0.06\right\} ,\left\{ 0.23,0.3,0.17\right\} \right\} $
and $\phi\left(0\right)=\left\{ \left\{ 0,0.001\pi,0.002\pi\right\} ,\left\{ \frac{\pi}{2}+0.001\pi,\frac{\pi}{2},\frac{\pi}{2}+0.0005\pi\right\} \right\} $.
The Hamiltonian is $H=2I$. From top to bottom, the functions $F$ within equations
\ref{eq:phidotnew} and \ref{eq:rhodotnew} for the plots are $F=\cos^{2}\left(\frac{\Delta\phi}{2}\right)$
and $F$ given by equation \ref{eq:spikeF} with $c=100$. Note that the standard deviation plots are logarithmic in scale on the y-axis, with the second such
plot being logarithmic in scale on both axes.}

\begin{tabular}{|c|c|c|c|}
\hline 
 & Probability vs Time & Phase Difference vs Time & Standard Deviation vs Time\tabularnewline
\hline
\hline 
\includegraphics[width=0.015\paperwidth]{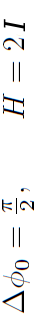} & \includegraphics[width=0.26\paperwidth]{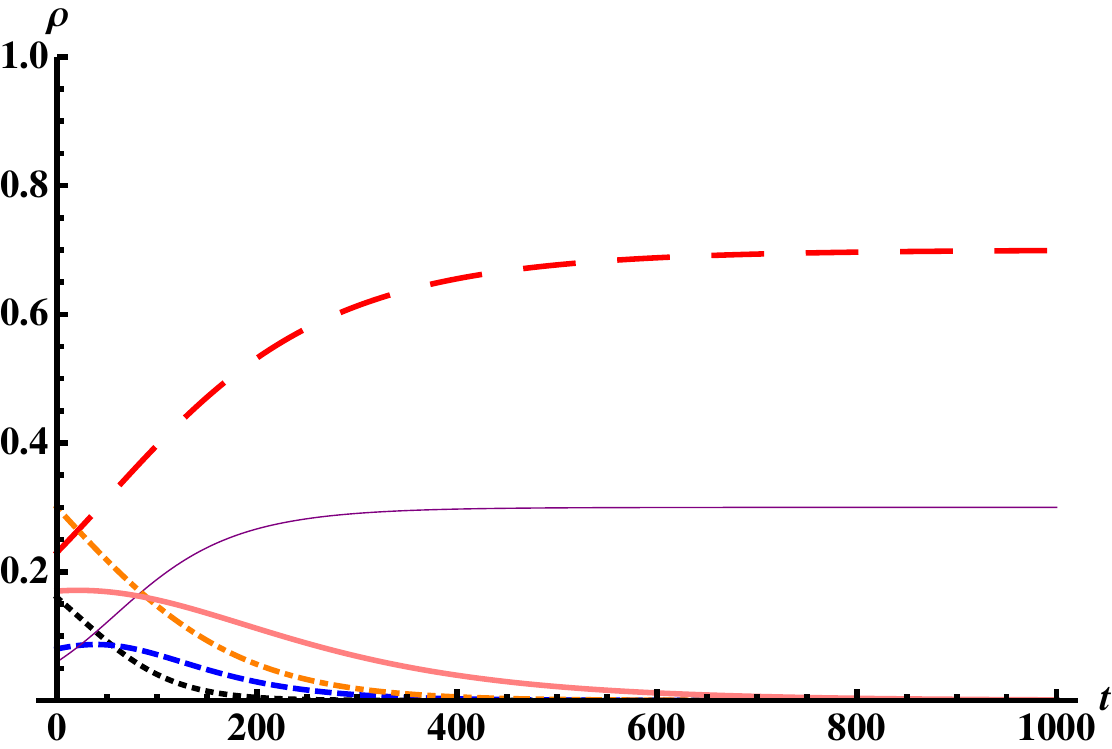} & \includegraphics[width=0.26\paperwidth]{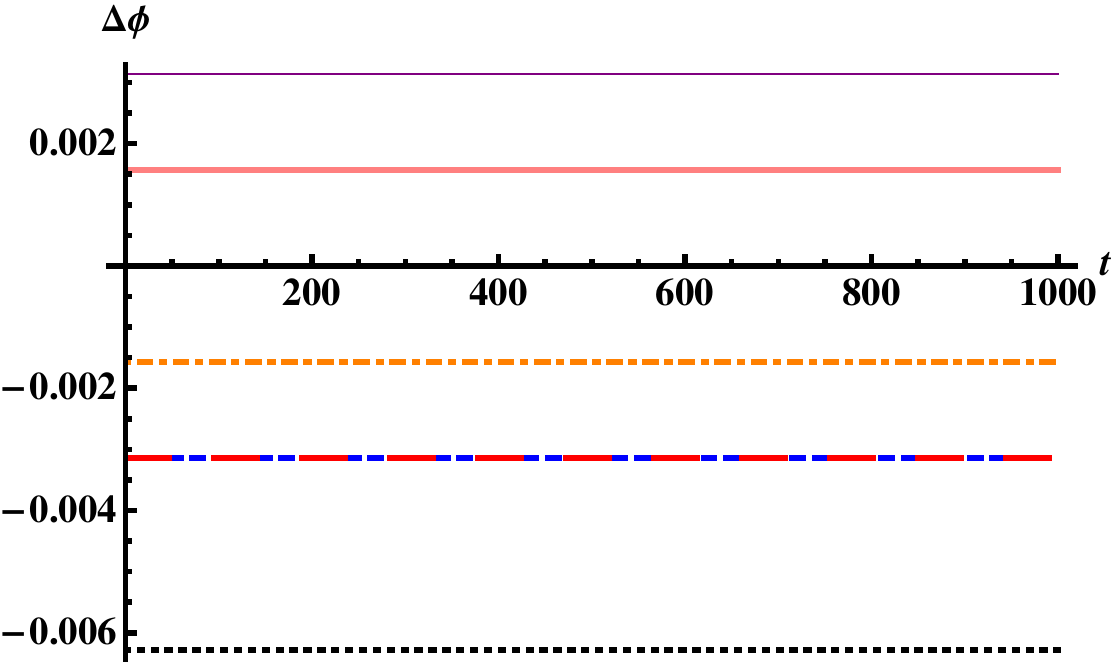} & \includegraphics[width=0.26\paperwidth]{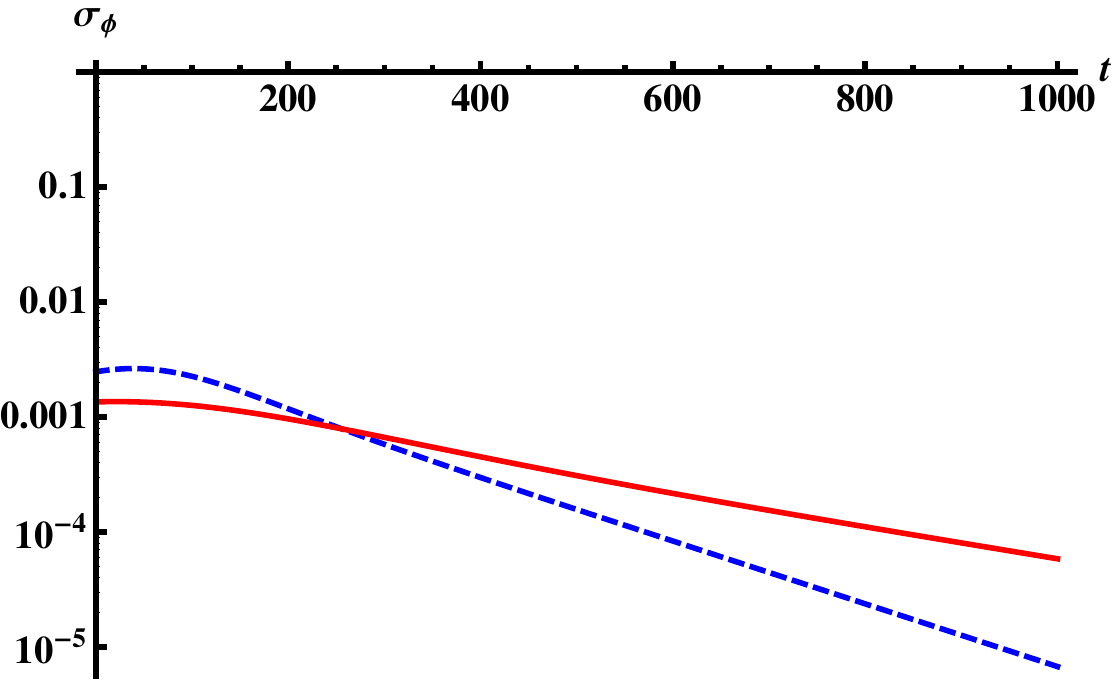}\tabularnewline
\hline 
\includegraphics[width=0.0173\paperwidth]{ALii2} & \includegraphics[width=0.26\paperwidth]{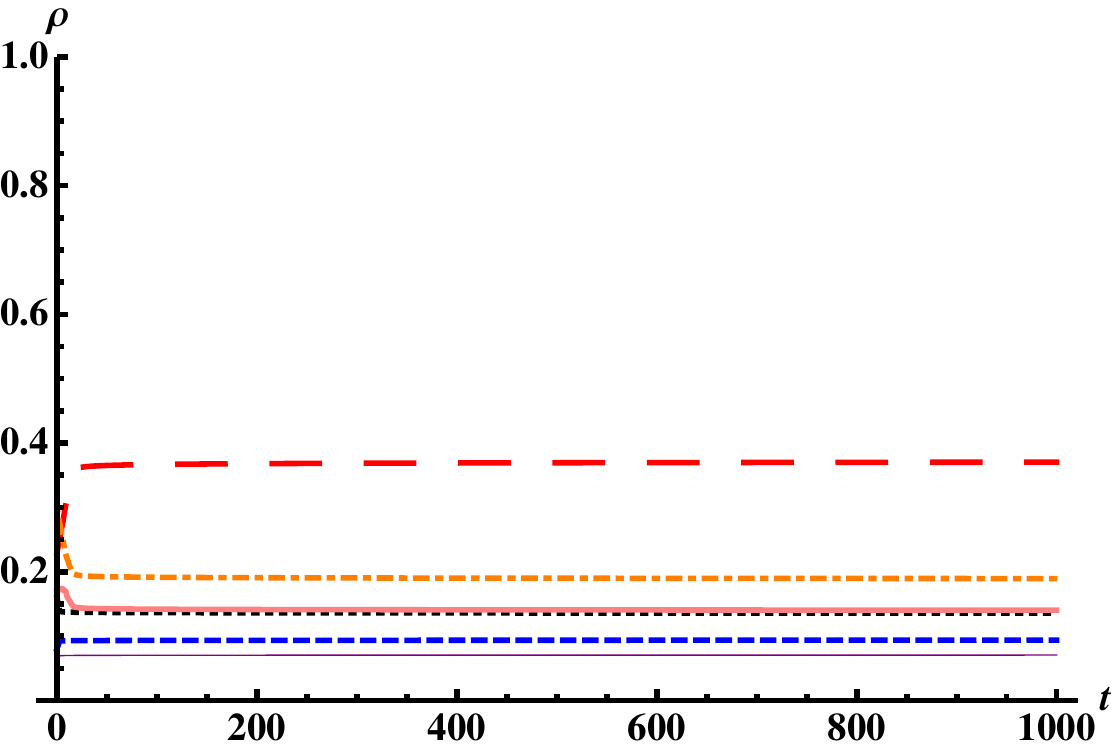} & \includegraphics[width=0.26\paperwidth]{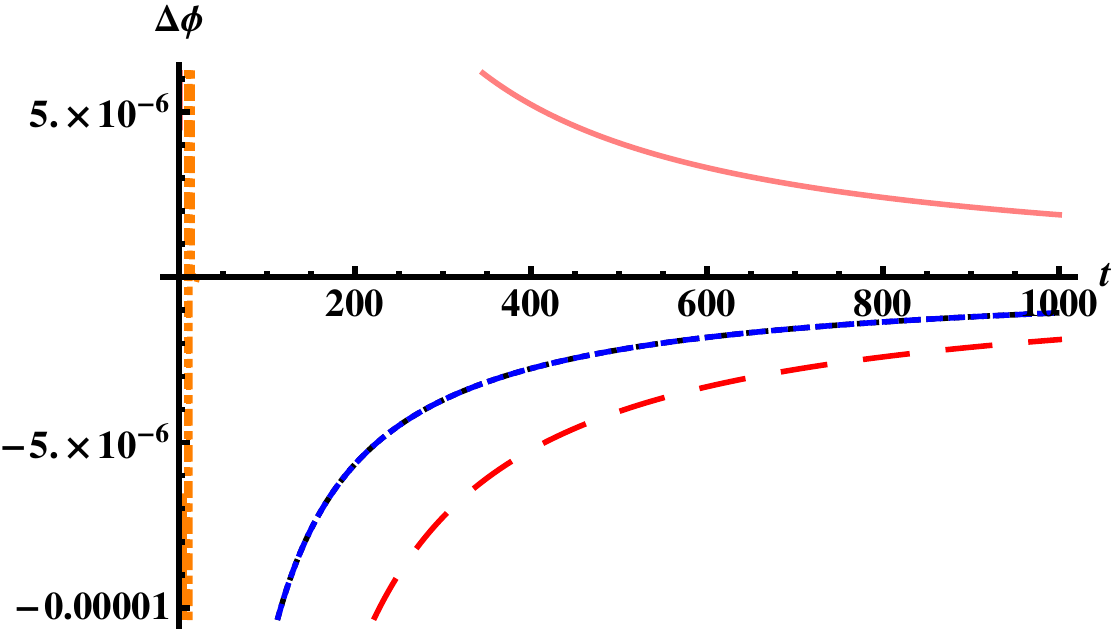} & \includegraphics[width=0.26\paperwidth]{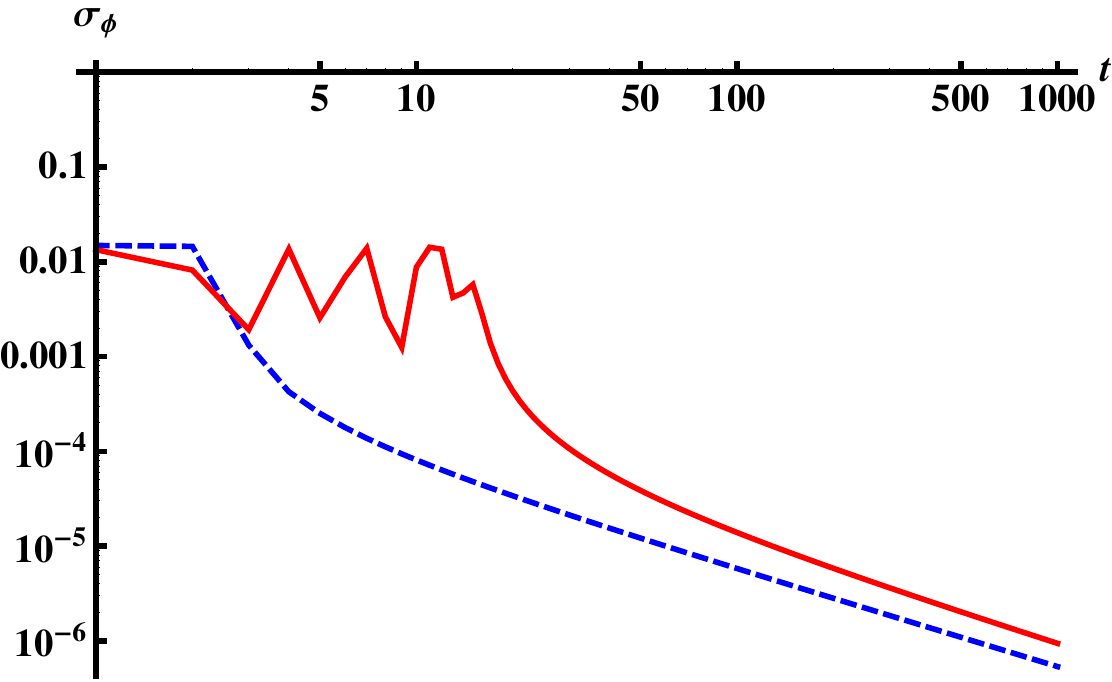}\tabularnewline
\hline
\end{tabular}
\end{table*}


\begin{thebibliography}{10}
\providecommand{\url}[1]{{#1}}
\providecommand{\urlprefix}{URL }
\expandafter\ifx\csname urlstyle\endcsname\relax
  \providecommand{\doi}[1]{DOI~\discretionary{}{}{}#1}\else
  \providecommand{\doi}{DOI~\discretionary{}{}{}\begingroup
  \urlstyle{rm}\Url}\fi

\bibitem{Bell:1964kc}
Bell, J.: {On the Einstein-Podolsky-Rosen paradox}.
\newblock Physics \textbf{1}, 195 (1964)

\bibitem{Liberati:2013xla}
Liberati, S.: {Tests of Lorentz invariance: a 2013 update}  (2013)

\bibitem{valentini2002signal}
Valentini, A.: Signal-locality in hidden-variables theories.
\newblock Physics Letters A \textbf{297}(5-6), 273--278 (2002)

\bibitem{Ade:2013ktc}
Ade, P., et~al.: {Planck 2013 results. I. Overview of products and scientific
  results}  (2013)

\bibitem{Horava:2009uw}
Horava, P.: {Quantum Gravity at a Lifshitz Point}.
\newblock Phys.Rev. \textbf{D79}, 084,008 (2009).
\newblock \doi{10.1103/PhysRevD.79.084008}

\bibitem{PhysRev.85.166}
Bohm, D.: A suggested interpretation of the quantum theory in terms of "hidden"
  variables. i.
\newblock Phys. Rev. \textbf{85}, 166--179 (1952).
\newblock \doi{10.1103/PhysRev.85.166}.
\newblock \urlprefix\url{http://link.aps.org/doi/10.1103/PhysRev.85.166}

\bibitem{PhysRev.85.180}
Bohm, D.: A suggested interpretation of the quantum theory in terms of "hidden"
  variables. ii.
\newblock Phys. Rev. \textbf{85}, 180--193 (1952).
\newblock \doi{10.1103/PhysRev.85.180}.
\newblock \urlprefix\url{http://link.aps.org/doi/10.1103/PhysRev.85.180}

\bibitem{quant-ph/0104067}
Valentini, A.: Hidden variables, statistical mechanics and the early universe.
\newblock Tech. Rep. Imperial/TP/0-01/16 (2001).
\newblock \urlprefix\url{http://arxiv.org/abs/quant-ph/0104067}

\bibitem{smolin2011real}
Smolin, L.: {A Real ensemble interpretation of quantum mechanics}.
\newblock Found.Phys. \textbf{42}, 1239--1261 (2012).
\newblock \doi{10.1007/s10701-012-9666-4}

\bibitem{Geshnizjani:2011dk}
Geshnizjani, G., Kinney, W.H., Dizgah, A.M.: {General conditions for
  scale-invariant perturbations in an expanding universe}.
\newblock JCAP \textbf{1111}, 049 (2011).
\newblock \doi{10.1088/1475-7516/2011/11/049}

\end{thebibliography}
\end{document}